\documentclass{aa}  

\usepackage{graphicx}
\usepackage{txfonts}
\usepackage{color}

\usepackage{bm}
\usepackage{media9}
\begin{document} 

   \title{Efficient magnetohydrodynamic modelling of the time-evolving corona by COCONUT}
   \author{H. P. Wang
          \inst{1}
          \and
          S. Poedts\inst{1,2}
          \and
          A. Lani\inst{1,3}
          \and 
          M. Brchnelova\inst{1}
          \and
          T. Baratashvili\inst{1}
          \and
          L. Linan\inst{1}
          \and
          F. Zhang\inst{4,5}
          \and
          D. W. Hou\inst{6}
          \and
          Y. H. Zhou\inst{1}
          }

   \institute{Centre for Mathematical Plasma-Astrophysics, Department of Mathematics, KU Leuven, Celestijnenlaan 200B,
   3001 Leuven, Belgium\\
  \email{Stefaan.Poedts@kuleuven.be}\\
  \email{andrea.lani@kuleuven.be}\\
  \email{haopeng.wang1@kuleuven.be}
\and 
Institute of Physics, University of Maria Curie-Skłodowska, ul.\ Radziszewskiego 10, 20-031 Lublin, Poland
\and
Von Karman Institute For Fluid Dynamics, Waterloosesteenweg 72, 1640 Sint-Genesius-Rode, Brussels, Belgium
\and
Institute of Theoretical Astrophysics, University of Oslo, PO Box 1029 Blindern, 0315 Oslo, Norway   
\and
Rosseland Centre for Solar Physics, University of Oslo, PO Box 1029 Blindern, 0315 Oslo, Norway    
\and
Department of Automation, University of Science and Technology of China, Hefei 230026, China 
}
  \abstract
   {Magnetohydrodynamic (MHD) solar corona models are critical in the Sun-to-Earth modelling chain and are the most complex and computationally intensive component. 
   Compared to quasi-steady-state corona models that are constrained by a time-invariant magnetogram over a Carrington rotation (CR) period, time-evolving corona models driven by time-varying photospheric magnetograms are more realistic and can maintain more useful information to accurately describe solar wind evolution and forecast coronal mass ejection propagation.} 
   {Implicit methods have significantly improved the efficiency of quasi-steady MHD coronal modelling. However, developing efficient time-evolving corona models to improve space weather forecasting is also important.
   This paper aims to demonstrate that time-evolving corona simulations can be performed efficiently and accurately using an implicit method with relatively large time steps, thus reducing the overall computational cost. We also evaluate differences between coronal structures captured by time-evolving and quasi-steady simulations over a CR period during solar minimum.}
   {We extended the quasi-steady COCONUT model, a global MHD corona model that uses implicit methods to select large time steps, into a time-evolving corona model. Specifically, we used a series of hourly updated photospheric magnetograms to drive the evolution of coronal structures from the solar surface to $25\; R_s$ during two CRs around the 2019 eclipse in an inertial coordinate system. At each time step, the inner-boundary magnetic field was temporal-interpolated and updated from adjacent observation-based magnetograms.  We compare the time-evolving and quasi-steady simulations to demonstrate that the differences in these two types of coronal modelling can be obvious even for a solar minimum. The relative differences in radial velocity and density can be over $15 \%$ and $25 \%$ at 20$\;R_s$ during one CR period. We also evaluated the impact of time steps on the simulation results. Using a time step of approximately 10 minutes balances efficiency and necessary numerical stability and accuracy for time-evolving corona simulations around solar minima. The chosen 10-minute time step significantly exceeds the Courant-Friedrichs-Lewy stability condition needed for explicit corona modelling, and the time-evolving COCONUT can thus simulate the coronal evolution during a full CR within only 9 hours (using 1080 CPU cores for 1.5M grid cells).}
   {The simulation results demonstrate that time-evolving MHD coronal simulations can be performed efficiently and accurately using an implicit method, offering a more realistic alternative to quasi-steady-state simulations. The fully implicit time-evolving corona model thus promises to simulate the time-evolving corona accurately in practical space weather forecasting.}
   {}

   \keywords{Sun: magnetohydrodynamics (MHD) --methods: numerical --Sun: corona}

   \maketitle
%

\section{Introduction}

Space weather refers to the variable physical conditions on the Sun and in the solar wind, the magnetosphere, and the ionosphere. It can influence the performance and reliability of space- and ground-based technological systems and affect human life and health. There is thus a need to develop advanced Sun-to-Earth model chains to understand the mechanisms of space weather and ultimately provide reliable space weather forecasts hours to days in advance \cite[e.g.][]{BAKER19987,Feng_2011Chinese,Feng_2013Chinese,Feng2020book,Koskinen2017}. Physics-based magnetohydrodynamic (MHD) modelling is the first principal method capable of self-consistently bridging large heliocentric distances from near the Sun to well beyond Earth's orbit  \cite[e.g.][]{Detman2005,Feng_2007,Feng_2014,Feng_2010,Feng_2011,Feng_2012,Feng2012,
FENG20141965,Feng_2017,Feng2020book,Gombosi2018,Hayashi2006,Lihuichao2018,LiHuichao2020,LUGAZ20111187,Mikic1999,
Nakamizoetal2009,RILEY20121,Shen_2021,TOTH2012870,Usmanov1993,Usmanov2003,WuShiTsan2015,Yang_2021,Zhou2012,Zhouyufen2017}. However, realistic MHD simulations of the solar-terrestrial system are complex, involving various physical phenomena across diverse spatiotemporal scales, and are very computationally intensive. We need to develop more efficient and reliable MHD models to improve space weather forecasting \cite[e.g.][and references therein]{Feng2020book,Owens2017}.

Since solar-terrestrial space involves diverse spatiotemporal scales and phenomena, using a single model to simulate the entire solar-terrestrial space is inefficient. Coupling different models, respectively dedicated to specific regions and physical problems, has become a preferred approach for establishing a space weather forecasting framework \cite[e.g.][]{Feng_2013Chinese,GOODRICH20041469,Hayashi_2021,ODSTRCIL20041311,Pomoell2018020,Poedts_2020,TOTH2012870}. In a coupled Sun-to-Earth modelling chain, observed photospheric magnetic fields serve as input data for the solar corona model. The solar corona model provides the inner heliospheric model's inner boundary conditions. The inner heliospheric model gives boundary information to the geomagnetic model. The solar corona model is crucial for determining the initialisation of the modelling chain. Thus, it is also a key factor affecting the simulation results of solar disturbance propagation and evolution \citep{Brchnelova_2022,Perri_2023}. Specifically, the solar wind speed increases from subsonic {and/or} sub-Alfv{\'e}nic to supersonic {and/or} super-Alfv{\'e}nic in the solar corona, and solar disturbances such as coronal mass ejections (CMEs) and solar proton events also propagate through this layer \citep{Feng2020book,Kuzma_2023}.

Though the solar corona is a crucial link in the Sun-to-Earth modelling chain and significantly impacts the ultimate effects of space weather, physics-based MHD corona models are also the most complex and computationally intensive component. Specifically, depending on the mesh resolution, even the currently state-of-the-art quasi-steady-state corona simulations take 10$\sim100\;$k CPU-hours to reach a quasi-steady state \citep{FengandLiu2019,Reville_2020}. In these simulations, time steps are limited to a few seconds due to the restriction of the Courant-Friedrichs-Lewy (CFL) stability condition. In contrast, the time steps limited by the CFL condition are typically of the order of 10 minutes for most MHD inner heliospheric models \citep{Detman2005,Hayashi2012}. As a result, MHD corona models require significantly more computational resources to remain synchronised with inner heliospheric models. Significant simplifications in solar corona modelling are frequently needed to improve efficiency. However, empirical solar corona models \citep{Arge2003ImprovedMF,Yangzicai2018} discard important information, and it has been demonstrated that even a relatively simple MHD model provides better forecasts \citep{Samara_2021}. Therefore, more efforts are required to establish more efficient and accurate MHD solar corona models.

\cite{Sokolov2021} solved 1D equations for the plasma motion and heat transport along fixed potential magnetic field lines together with the Alfv{\'e}nic wave propagation in the low coronal region and interfaced this threaded-field-line model with the full MHD global corona model at $1.1\;R_s$. This enabled the updated Alfv{\'e}n wave solar atmosphere model (AWSoM) model, AWSoM-R (AWSoM-realtime), to achieve a faster-than-real-time performance on approximately 200 CPU cores. In contrast, the original AWSoM model required 1000-2000 CPU cores to reach a similar computing time \citep{Jin_2017}. More generally, implicit temporal discretisation strategies can be used to overcome the limitation imposed by the CFL condition, thereby improving the overall computational efficiency by utilising a larger time step.

Recently, several successful attempts have been made to increase the efficiency of MHD corona models by using implicit solvers. For instance, \cite{WANG2019181} achieved speedup ratios of 31.27 and 28.05 in MHD solar corona simulations with an effective matrix-free implicit scheme. \cite{Feng_2021} and \cite{Wang_2022,Wang2022_CJG} further improved the implicit method and developed an efficient parallel lower-upper symmetric Gauss-Seidel solver, improving computational efficiency even with a plasma $\beta$ smaller than $10^{-4}$.  COolfluid COroNal UnsTructured (COCONUT), a novel MHD solar corona model based on the Computational Object-Oriented Libraries for Fluid Dynamics (COOLFluiD)\footnote{\url{https://github.com/andrealani/COOLFluiD.git}}, gained a speed up of 35 compared to the state-of-the-art time-explicit wind-predict model \citep{Perri2018SimulationsOS} while achieving the same level of accuracy in steady-state simulations \citep{Perri_2022,Perri_2023}. In addition, CME simulations \citep{guo2023,Linan_2023} demonstrate that the  COCONUT model has the potential to be faster than the explicit MHD SC models in time-dependent simulations while maintaining time-accurate using a relatively large time step.
\cite{WangSubmitted} proposed an efficient time-accurate implicit MHD model for the solar corona and CMEs, which is capable of timely and accurately simulating time-varying events in the solar corona, even with low plasma beta. The fully implicit scheme of the block-adaptive tree solarwind Roe-type upwind scheme (BATS-R-US) code \citep{TOTH2012870}, the explicit scheme of which was used by AWSoM corona model, can also produce speedup ratios of the order of 10-20 compared to the explicit version in simulations of some geophysical applications \citep{TOTH2006722,TOTH2008}. 

However, it is still necessary to further develop coronal modelling to capture time-evolving solar coronal structures more accurately, as current models frequently assume a (quasi-)steady corona during one Carrington rotation (CR) and use a time-invariant photospheric magnetic field as the inner boundary. The (quasi-)steady simplification differs from the reality that the solar coronal structure evolves \citep{Owens2017} and thus leads to discrepancies between the simulation results and observations of the solar coronal structures \citep{Cash_2015,Reville_2020}. Time-evolving corona models typically driven by hundreds of time-evolving observed photospheric magnetograms can capture time-evolving coronal structures with higher fidelity \citep{Feng_2023,Yang2012} and thus may also improve solar wind and CME modelling \citep{Lionello_2023}.

For instance, \cite{Detman2005}, \cite{Linker_2016}, and \cite{Merkin2016} used a time series of photospheric magnetograms to drive an empirical source surface current sheet corona model to provide time-evolving lower boundary conditions for the inner heliosphere MHD solar wind model. These hybrid Sun-to-Earth modelling systems are efficient for real-time operations but fail to produce coronal transient events or model closed loops that rise beyond the source surface (usually between 2.0 and 2.5 $R_s$). \cite{Hoeksema2020} and \cite{Hayashi_2021} employed the electric field derived at the Sun's photosphere from a sequence of vector magnetogram and Doppler velocity measurements \citep{Fisher_2020} to drive a magneto-frictional (MF) model, and then used the MF model to produce time-evolving boundary magnetic fields at 1.15$\;R_s$ for their global coronal-heliospheric MHD (GHM) model. Although the MF model can be numerically stable in generating time-dependent 3D coronal magnetic structures, it failed to provide the required initial dynamic states of the plasma in the low atmosphere. To ensure the boundary plasma quantities at 1.15$\;R_s$ evolve consistently with both the variations of the magnetic field specified from the MF model and the governing MHD equations of the GHM model, they employed the projected normal characteristics method \cite[e.g.][]{Sauerwein_1966} at the interface boundary.  \cite{Feng_2023} and \cite{Yang2012} specified the tangential boundary electric field to make the flux evolution match the changes of the observed radial magnetic field and employed the projected normal characteristic method to make boundary conditions self-consistent. \cite{Feng_2012}, \cite{Lionello_2023}, and \cite{Mason_2023} employed the surface flux transport model \citep{DeVore1984,Schrijver2003} to obtain input maps that incorporate magnetic flux emergence and surface flows for their MHD corona models.

Another time-evolving corona model, magnetohydrodynamic algorithm outside a sphere \citep[MAS;][]{Lionello_2023,Mason_2023}, adopts a semi-implicit approach in which only some source terms are treated implicitly, requiring the time step to be selected according to the limitations imposed by the explicitly treated terms. \cite{Feng_2023} used the fully implicit method \citep{Feng_2021,Wang_2022} only at the six-layer grid close to the solar surface in the radial direction and reported that it helped avoid the occurrence of negative density and pressure caused by the strong magnetic field near the Sun. The remaining models still use explicit methods. 

We have extended the quasi-steady-state COCONUT to a time-evolving corona model and uses cubic Hermite interpolation to derive the time-evolving magnetograms at each physical time step. This extension demonstrates that the time-evolving corona model can achieve significantly better efficiency and numerical stability by adopting fully implicit algorithms appropriately. 

For robust solar corona modelling, ensuring the positivity-preserving (PP) property of thermal pressure and density in MHD simulations is vital. For example, a self-adjusting PP reconstruction method, initially proposed by \cite{Balsara2012} for solving hydrodynamic and MHD equations, was implemented in solar corona simulations by \cite{Feng_2017}. It was applied to conservative variables via a flattener function defined according to the flow's rarefactive and compressive motions. Furthermore, \cite{Feng_2021} and \cite{Wang_2022} extended this method to primitive variables and implemented it in implicit MHD corona models. Additionally, some PP Harten-Lax-van Leer (HLL) Riemann solvers \citep{Wu2019} have been used to design PP MHD corona models \citep{Feng_2021,Wang_2022}. It has been shown that the approximate reconstruction method and flux solver are beneficial to maintaining the PP property of MHD models. In this study, under the cell-centred finite-volume framework of COCONUT, we implemented the Venkatakrishnan limiter \citep{VENKATAKRISHNAN1993} in the reconstruction formula of primitive variables as before to control spatial oscillation. Furthermore, we designed an HLL Riemann solver with a self-adjustable dissipation term to accommodate low- and high-speed flows.

\cite{brchnelova2023assessing} improved the PP property of COCONUT by manipulating the inner-boundary density according to the local Alfv{\'e}nic velocity. In this method, the active region density was carefully increased to avoid an abnormally high Alfv{\'e}nic speed when the local inner-boundary Alfv{\'e}nic velocity exceeds a prescribed maximum Alfv{\'e}nic speed. This method can prevent some non-physical negative thermal pressures and very high-speed streams developed in the domain above the active regions \citep{Kuzma_2023}. Consequently, the performance, in terms of both convergence and physical accuracy, can be improved. To enhance the PP property of this time-evolving MHD corona model, we further extended this method to all the computational domains and applied a smooth hyperbolic tangent function to gradually increase the plasma density when the local Alfv{\'e}nic speed exceeds 2000$\;\rm {km\big/ S}$. In the future, we can take some extra measures, such as applying the self-adjusting PP reconstruction method \citep{Feng_2021,Wang_2022} and incorporating a mass flux limitation \citep{Hayashi_2005,Yang2012}, to enhance the PP property of COCONUT further.

Moreover, reducing the magnetic field discretisation error is also helpful in maintaining the PP property of MHD models in low $\beta$ regions. Considering that $\left(\mathbf{B}+\epsilon~\mathbf{B}\right)^2-\mathbf{B}^2\equiv 2~\epsilon~\mathbf{B}^2+\epsilon^2~\mathbf{B}^2$, with $\epsilon~\mathbf{B}$ denoting the magnetic field discretisation error, the magnetic pressure discretisation error can be comparable to thermal pressure in low $\beta$ (the ratio of the thermal pressure to the magnetic pressure) regions, and non-physical negative thermal pressure is prone to appear when deriving thermal pressure from energy density. Some researchers try to decrease the magnetic field discretisation error by adopting fine meshes near the solar surface to avoid such an undesirable situation. In AWSoM, a spherical grid ranging from 1 $R_s$ to 24 $R_s$ is used for the coronal component. The grid is highly refined in the radial direction near the Sun, with the smallest radial grid spacing being approximately 700 $\rm km$ and the total number of cells being 29.7 M in \cite{van_der_Holst_2022}. For the MAS model, a non-uniform spherical grid consisting of 27.3 M cells and covering a radial range from 1 $R_s$ to 30 $R_s$ is used in \cite{Caplan_2017}. It is highly non-uniform in the radial direction, and the smallest cell size in the radial direction is 340 $\rm km$. This paper adopts the unstructured geodesic mesh \citep{Brchnelova2022,Perri_2022},  which consists of approximately 1.5 M cells. The grid covers a radial range from 1 $R_s$ to 25 $R_s$ and gradually stretches in the radial direction with an initial grid spacing of about 170 $\rm Km$ at the solar surface. 

The paper is organised as follows. In Sect. \ref{sec:Governing equations and computational grid system} we introduce the governing equations, grid system, and initial conditions adopted in the solar corona simulations. In Sect. \ref{sediscretisationof the novel MHD Model} and Appendix \ref{HLLsad}, the numerical formulation of the time-evolving corona simulations is described in detail. These sections mainly describe the discretisation of the MHD equations, the implementation of time-evolving boundary conditions, the HLL Riemann solver with a self-adjustable dissipation term, and the PP measures used to enhance the corona model's numerical stability. We demonstrate the simulation results in Sect.~\ref{sec:Numerical Results}: the evolution of the corona during two CR periods around the 2019 eclipse simulated by the time-evolving COCONUT is demonstrated, a comparison of the 2019 eclipse simulations performed by both the time-evolving COCONUT and its quasi-steady-state version is illustrated, and the simulation results calculated with different time-step sizes are compared. In Sect. \ref{sec:Conclusion} we summarise the main features of the efficient, fully implicit, time-evolving corona model and give some concluding remarks.

\section{Governing equations and grid system}\label{sec:Governing equations and computational grid system}
This section mainly describes the governing equations, grid system, and initial conditions for simulating quasi-steady-state corona and time-evolving corona simulations.

\subsection{The governing equations}\label{The governing equations}
In this study
we developed the time-evolving version of COCONUT for time-evolving corona simulations.  First, we initiated a quasi-steady-state corona simulation constrained by a fixed magnetogram to get the background corona. Once the steady-state simulation converged, we drove the subsequent evolution of the dynamic corona by a series of time-evolving photospheric magnetograms. 

The governing MHD equations are calculated in the heliocentric inertial (HCI) coordinate system \citep{Burlaga1984MHDPI,FRANZ2002217}:\begin{equation}\label{MHDinsolarwind}
\frac{\partial \mathbf{U}}{\partial t}+\nabla \cdot \mathbf{F}\left(\mathbf{U}\right)=\mathbf{S}\left(\mathbf{U},\nabla \mathbf{U}\right).\
\end{equation}
Here $\mathbf{U}=\left(\rho, \rho \mathbf{v}, \mathbf{B}, E, \psi\right)^T$ denotes the conservative variable vector, $\nabla \mathbf{U}$ corresponds to the spatial derivative of $\mathbf{U}$,  the inviscid flux vector $\mathbf{F}\left(\mathbf{U}\right)$ is\begin{equation*}
    \mathbf{F}\left(\mathbf{U}\right)=
                     \begin{pmatrix}
\rho \mathbf{v}  \\
\rho\mathbf{v}\mathbf{v}+p_{T}\mathbf{I}-\mathbf{B}\mathbf{B}\\
\mathbf{vB}-\mathbf{Bv}+\psi\mathbf{I}\\
\left(E+p_{T}\right)\mathbf{v}-\mathbf{B}\left(\mathbf{v}\cdot\mathbf{B}\right)\\
V_{\rm ref}^2 \mathbf{B}
                     \end{pmatrix},\
\end{equation*}
and $\mathbf{S}\left(\mathbf{U},\nabla \mathbf{U}\right)=\mathbf{S}_{\rm gra}+\mathbf{S}_{\rm heat}$ represents the source term vector corresponding to the gravitational force and the heating source terms defined as follows:
\begin{equation}\label{Svolumheat}
\mathbf{S}_{\rm gra}=-\frac{\rho G M_s}{\left|\mathbf{r}\right|^3}
                        \begin{pmatrix}
                        0\\
                        \mathbf{r}\\
                        \mathbf{0}\\
                        \mathbf{r}\cdot\mathbf{v} \\
                        0
                        \end{pmatrix}, \quad
\mathbf{S}_{\rm heat}=\begin{pmatrix}
                        0\\
                        \mathbf{0}\\
                        \mathbf{0}\\
                        -\nabla \cdot \mathbf{q}+Q_{rad}+Q_{H} \\
                        0
                        \end{pmatrix}.\
\end{equation}

In these formulations mentioned above, 
$\mathbf{B}=\left(B_x,B_y,B_z\right)$ and $\mathbf{v}=\left(u,v,w\right)$ denote the magnetic field and velocity in Cartesian coordinate system, $E=\frac{p}{\gamma-1}+\frac{1}{2}\rho\mathbf{v}^{2}+\frac{1}{2}\mathbf{B}^{2}$ means the total energy density with the adiabatic index $\gamma=\frac{5}{3}$, $p_{T}=p+\frac{\mathbf{B}^2}{2}$ is the total pressure, $\rho$ and $p$ represent the density and thermal pressure of the plasma, $\mathbf{r}$ is the position vector, $r=\left|\mathbf{r}\right|$ denotes the heliocentric distance, and $t$ represents the time. For convenience of description, in the definition of the magnetic field, a factor of $\frac{1}{\sqrt{\mu _0}}$ is absorbed with $\mu _0 = 4 \times 10^{-7} \pi ~ \rm H  ~ m^{-1}$ denoting the magnetic permeability. As usual, $G$ means the universal gravitational constant, $M_s$ means the mass of the Sun, and $G M_s=1.327474512 \times 10^{20} ~\rm m^3~s^{-2}$.  The thermal pressure of the plasma is defined as $p= \Re \rho T$, where $T$ is the temperature of the bulk plasma, $\Re=1.299 \times 10^4 \rm m^2~s^{-2}~K^{-1}$ denotes the gas constant and is calculate by $\Re=\frac{2*k_B}{m_{\rm cor}*m_H}$ with $k_B=1.3806503 \times 10^{-23}\rm ~J~K^{-1}$ denoting the Boltzmann constant, the molecular weight setting to $m_{\rm cor}=1.27$ \citep{Aschwanden2005} and $m_H=1.67262158 \times 10^{-27}~ \rm kg$ representing the mass of hydrogen. We adopted the hyperbolic generalised Lagrange multiplier method \citep{Dedner2002JCoPh,YALIM20116136} to constrain the divergence error; $\psi$ and $V_{\rm ref}$ denote the Lagrange multiplier and the propagation speed of the numerical divergence error. In the energy source term, $Q_{H}$, $Q_{rad}$, and $-\nabla \cdot \mathbf{q}$ are the coronal heating, radiation loss, and thermal conduction, respectively.

\begin{figure}[htpb]
\begin{center}
  \vspace*{0.01\textwidth}
    \includegraphics[width=\linewidth,trim=1 1 1 1, clip]{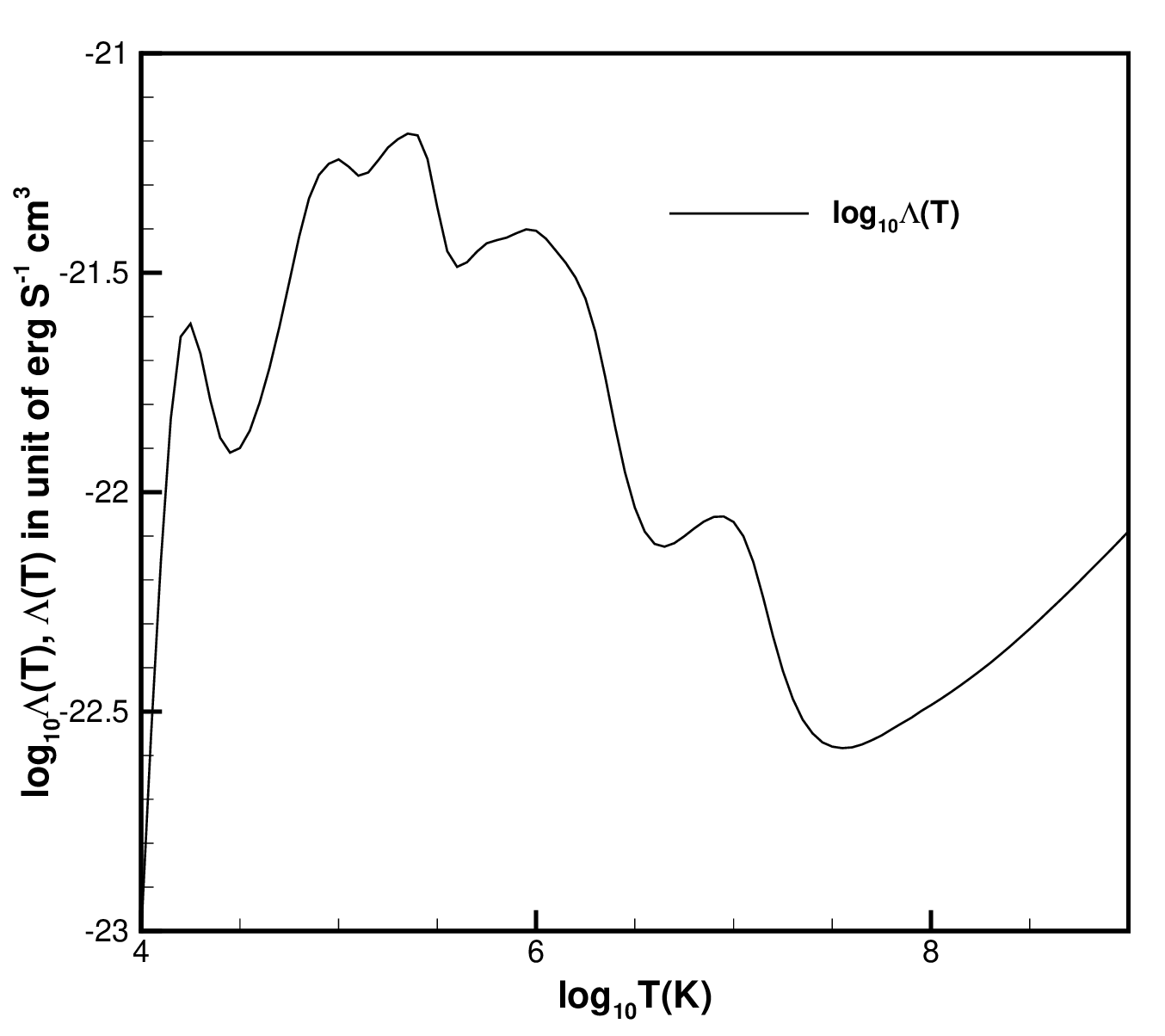}
\end{center}
\caption{Radiative cooling curve profile derived from version 9 of the CHIANTI atomic database  \citep{Dere_2019}. The horizontal axis denotes the decadic logarithm of temperature, and the vertical axis the logarithm of the radiative cooling curve function value. The temperature and radiative cooling curve function units are $\rm K$ and $\rm erg~S^{-1}~cm^3$, respectively.}\label{RadiativeLosCurchiant9}
\end{figure}

As done in \cite{Baratashvili}, the heat flux, $\mathbf{q,}$ included in $\mathbf{S}_{\rm heat}$ is defined in a Spitzer or collisionless form according to the radial distance \citep{Hollweg1978,Mikic1999}:
\begin{equation}\label{heatflux}
\mathbf{q}=\left\{\begin{array}{c}
-\xi T^{5 / 2}(\hat{\mathbf{b}} \cdot \nabla T) \hat{\mathbf{b}},  \text { if } 1 \leq r \leq 10~ R_{s} \\
\alpha n_{e} k_B T \mathbf{v},  \text { if } r>10~ R_{s}
\end{array}\right..\
\end{equation}
Here $\hat{\mathbf{b}}=\frac{\mathbf{B}}{\left|\mathbf{B}\right|}$, $\xi =9.0 \times 10^{-12} \rm J m^{-1} s^{-1} K^{-\frac{7}{2}}$, $\alpha$ is set to $\frac{3}{2}$ \citep{Lionello_2008} and $n_{e}$ is the electron number density. By assuming the radiative loss to be optically thin \citep{Rosner1978,Zhou2021}, the radiative term is\begin{equation}\label{Radiationloss}
Q_{rad}=-n_e n_p \Lambda\left(T\right),\
\end{equation}
where the proton number density, $n_p$, is assumed to be equal to the electric number density, $n_e$, for the hydrogen plasma. $\Lambda\left(T\right)$ is a temperature-dependent radiative cooling curve function. Similar to \cite{van_der_Holst_2014}, $\Lambda\left(T\right)$ in this paper is derived from version 9 of CHIANTI \citep{Dere_2019}, an atomic database for emission lines. The profile of $\log\left(\Lambda\left(T\right)\right)$ along $\log\left(T\right)$, with $\Lambda\left(T\right)$ and $T$ in unite of $\rm erg~S^{-1}~cm^3$ and $\rm K$, is shown in Fig.~\ref{RadiativeLosCurchiant9}.
As did in \cite{Xia_2011}, we set $\Lambda\left(T\right)$ to zero when $T<2 \times 10^4\;$K, which means the plasma has become optically thick and is no longer fully ionized.
We adopted the following empirical heating source term discussed and recommended in {\cite{Baratashvili}. It is proportional to the local magnetic field strength \citep{Mok_2005} while also exhibiting an exponential decay in the radial direction \citep{Downs2010}:
\begin{equation}\label{Coronalheating}
Q_{H}=H_0 \cdot \left|\mathbf{B}\right|\cdot e^{-\frac{r-R_s}{\lambda}},\
\end{equation}
where $H_0=4\times 10^{-2} ~\rm J~m^{-3}~S^{-1}~T^{-1}$ and $\lambda=0.7 R_s$.

The variables $\mathbf{r}$, $\rho$, $\mathbf{v}$, $p$, $\mathbf{B,}$ and $t$ are normalized by $R_s$, $\rho_s$, $V_{a,s}$, $\rho_s V_{a,s}^2$, $B_s$, and $\frac{R_s}{V_{a,s}}$, respectively. Here $R_s=6.95 \times 10^8 ~ \rm m$ is the solar radius and $\rho_s=1.67\times 10^{-13} ~ \rm Kg~m^{-3}$, $B_s=2.2 \times 10^{-4}~\rm T,$ and $V_{a,s}=\frac{B_s}{\rho_s ^{0.5}}$ denote the characteristic plasma density, magnetic field strength, and Alfv{\'e}nic velocity at the solar surface, respectively.

\subsection{Computational grid system and initialisations}\label{Thecomputationalgridsystemandinitialvaluesetup}
The governing equations are solved using the cell-centred finite-volume method. The computational domain is a spherical shell ranging from 1.01 to 25 $R_s$ and discreted into an unstructured sixth-level subdivided geodesic mesh \citep{Brchnelova2022}, which includes 1495040 non-overlapped pyramid cells, each consisting of 2 triangular faces and 3 quadrilateral faces. There are 73 layers of gradually stretched cells in the radial direction, each containing 20480 cells. 

The observed GONG-zqs {(zeropoint-corrected) hourly-updated synoptic maps of the photospheric magnetic field} provide inner-boundary conditions of the magnetic field \citep{LiHuichao2021,Perri_2023}. Considering that the original photospheric magnetic field strength can be hundreds of Gauss near active regions and may cause non-physical negative temperatures or pressures, and the magnetic maps reconstructed by a potential-field (PF) solver of 20-order spherical harmonic expansion are already fully sufficient for space weather forecasting \citep{Kuzma_2023}, we used a PF solver that filters out spherical harmonic expansions above the 20th order to eliminate very small-scale structures from the original magnetograms. These preprocessed photospheric magnetograms constrain the corona simulations in COCONUT \citep{Kuzma_2023,Perri_2022,Perri_2023}. Meanwhile, the temperature and plasma density at the solar surface are set to $1.5 \times 10^6 ~\rm K$ and $1.67\times 10^{-13} ~ \rm Kg~m^{-3}$, respectively.

\section{Numerical methods}\label{sediscretisationof the novel MHD Model}
This section introduces the discretisation of the MHD equations and the implementation of inner-boundary conditions.

\subsection{Spatial discretisation and temporal integration}\label{discretization}
Godunov's method is used in COCONUT to advance cell-averaged solutions in time by solving a Riemann problem at each cell interface \citep{Godunov1959Adifference}.
By integrating Eq. (\ref{MHDinsolarwind}) over the prismatic cell $\rm{cell}_i$ and applying Gauss's theorem to calculate the volume integral of the flux divergence, we obtained the following discretized equations:
\begin{equation}\label{MHDequationdiscrization}
V_{i}\frac{\partial \mathbf{U}_{i}}{\partial t}=-\oint_{\partial V_{i}}\mathbf{F}\cdot \mathbf{n} d \Gamma+V_{i}\mathbf{S}_{i},\
\end{equation}
where $\oint_{\partial V_{i}}\mathbf{F}\cdot \mathbf{n} d \Gamma=\sum\limits_{j=1}^{5}\mathbf{F}_{ij}\cdot \mathbf{n}_{ij} \Gamma_{ij}$ and $\mathbf{S}_i=\mathbf{S}_{{\rm gra},i}+\mathbf{S}_{{\rm heat},i}$. $\mathbf{U}_{i}$ and $\mathbf{S}_{i}$ denote the cell-averaged solution variables and source terms in ${\rm cell}_i$, $V_{i}$ is the volume of ${\rm cell}_i$, $\Gamma_{ij}$ means the interface shared by ${\rm cell}_i$ and its neighbouring ${\rm cell}_j$, and also denotes the area of this interface, $\mathbf{n}_{ij}=\left(n_{x,ij},n_{y,ij},n_{z,ij}\right)$ is the unit normal vector of $\Gamma_{ij}$, oriented from ${\rm cell}_i$ to ${\rm cell}_j$, and $\mathbf{F}_{ij}\cdot \mathbf{n}_{ij}$ is described as $\mathbf{F}_{n,ij}$, denoting the inviscid flux along normal direction of $\Gamma_{ij}$. 

We employed an HLL Riemann solver \citep{EINFELDT1991273} with a self-adjustable dissipation term, as described in Appendix \ref{HLLsad}, to compute the inviscid flux $\mathbf{F}_{n,ij}$. The cell-averaged source terms $\mathbf{S}_{{\rm gra},i}$ and $Q_{rad,i}$ and $Q_{H,i}$ were calculated by substituting the corresponding variables at the centroid of ${\rm cell}_i$ into formulations of $\mathbf{S}_{\rm gra}$ and $Q_{rad}$ and $Q_{H}$. $\left(\nabla\cdot\mathbf{q}\right)_i$ was calculated using the Green-Gauss theorem as follows: $$\left(\nabla\cdot\mathbf{q}\right)_i=\frac{1}{V_{i}} \sum\limits_{j=1}^{5}  \mathbf{q}_{ij} \cdot \mathbf{n}_{ij} \Gamma_{ij},$$ 
where $\mathbf{q}_{ij}=\mathbf{q}\left(T_{ij},\left(\nabla T\right)_{ij},\mathbf{U}_{ij}\right)$ represent the heat flux through $\Gamma_{ij}$. In this formulation, $T_{ij}$ is derived from the equation of state $T_{ij}=\frac{p_{ij}}{\Re \rho_{ij}}$.
The plasma states at the cell surface $\Gamma_{ij}$ are required for the calculation of the inviscid flux $\mathbf{F}_{n,ij}$ and heat flux $\mathbf{q}_{ij} \cdot \mathbf{n}_{ij}$. We used a second-order accurate reconstruction method to calculate the piecewise linear polynomial of the primitive variables:
\begin{equation}\label{FlowfieldbyLSQ}
P_i(\mathbf{x})=\left.P\right|_i+\phi_i\left.\left(\nabla P\right)\right|_i\cdot\left(\mathbf{x}-\mathbf{x}_i\right)
,\end{equation}
where $P \in \{\rho,u,v,w,B_x,B_y,B_z,p, \psi\}$, $\left.P\right|_i$ is the corresponding primitive variables at the centroid of ${\rm cell}_i$ $\mathbf{x}_i$, $\left.\left(\nabla P\right)\right|_i=\left.\left(\frac{\partial P}{\partial x},\frac{\partial P}{\partial y},\frac{\partial P}{\partial z}\right)\right|_i$ is the derivative of $P$ at $\mathbf{x}_i$ calculated by least square method \citep{BARTH1993,Lani2008AnOO}, and $\phi_i$ is the Venkatakrishnan limiter \citep{VENKATAKRISHNAN1993} used to control spatial oscillation.

In the quasi-steady corona simulations, we applied a backward Euler temporal integration to Eq.~(\ref{MHDequationdiscrization}) and obtained the following equation:\begin{equation}\label{implicitbackwardEuler}
V_{i}\frac{\Delta \mathbf{U}^n_{i}}{\Delta t}+\mathbf{R}_{i}\left(\mathbf{U}^{n+1}\right)=\mathbf{0}.\
\end{equation}
The superscripts `$^{n}$' and `$^{n+1}$' denote the time level, $\mathbf{R}_i\left(\mathbf{U}^{n+1}\right)=\sum\limits_{j=1}^{5}\mathbf{F}_{n,ij}\left(\mathbf{U}^{n+1}_{L},\mathbf{U}^{n+1}_{R}\right)\Gamma_{ij}-V_{i}\mathbf{S}^{n+1}_i$ means the residual operator over ${\rm cell}_i$ at the $(n+1)$-th time level, $\Delta \mathbf{U}_i^{n}=\mathbf{U}_i^{n+1}-\mathbf{U}_i^{n}$ represents the solution increment between the $n$-th and $(n+1)$-th time levels, and $\Delta t=t^{n+1}-t^n$ is a user-defined time increment.  In the quasi-steady corona simulations, the time variable $t$ does not refer to a physical time but a relaxation time used to get a quasi-steady coronal structure.

After reaching a steady-state coronal structure, we utilised a series of time-varying magnetograms to drive the following evolution of the dynamic corona. Along with spatial accuracy, temporal accuracy is also crucial in time-evolving coronal simulations. To enhance the temporal accuracy of the implicit solver with time step lengths exceeding the CFL condition, we adopted the second-order accurate backward differentiation formula (BDF2) for temporal integration. Newton iterations are used within each time step of the implicit algorithm to update the intermediate state until the $L_2$ norm of the differences in state variables between two consecutive iterations decreases to $10^{-3}$, or until ten iterations are reached.

To enhance the PP property of COCONUT, the density updated during the Newton iterations is appropriately adjusted according to the local Alfv{\'e}nic velocity. This manipulation ensures that the local Alfv{\'e}nic velocities calculated from the updated intermediate coronal state remain within a reasonable range, thereby enhancing the PP property of COCONUT. In this adjustment, we applied a smooth hyperbolic tangent function to gradually increase the plasma density when the local Alfv{\'e}nic speed reaches a prescribed maximum Alfv{\'e}nic speed, $V_{A,max}$, as follows:\begin{equation}\label{PPRhoInner}
\rho=\Upsilon_{\rho}\frac{\mathbf{B}^2}{V^2_{A,max}}+\left(1-\Upsilon_{\rho}\right)\rho_{o}
,\end{equation}
where $\Upsilon_{\rho}=0.5+0.5\cdot \tanh\left(\frac{V_A-V_{A,max}}{V_{fac}}\cdot \pi\right)$ with $V_A=\frac{\left|\mathbf{B}\right|}{\rho_o^{0.5}}$, $V_{A,max}=2000~\rm {Km~ S^{-1}}$, and $V_{fac}=2~\rm {Km~ S^{-1}}$. Here, the subscript `$_{o}$' on $\rho$ refers to the density updated in the Newton iteration without adjustment.

\subsection{Implementation of boundary conditions}\label{sec:Brief overview of the application of UBC in the inner boundary}
We first performed a quasi-steady-state coronal simulation constrained by a time-invariant magnetogram \citep{Baratashvili}, and then evolved the magnetograms to drive the following time-evolving coronal simulation.

We adopted one layer of ghost cells near the inner and outer boundaries \citep{Lani2008AnOO}. The primitive variables in ghost cells near the inner boundary are defined as
\begin{equation}
P_G=2 P_{BC}-P_{inner},\
\end{equation}
where the subscripts `$_G$' and `$_{inner}$' denote the corresponding variables at the ghost cell's and the nearest inner cell's centroid, respectively. In this study we set $\psi_{BC}=0$ \citep{Perri_2022}, and $\rho_{BC}$ and $p_{BC}$ were defined following \cite{brchnelova2023assessing} to improve the PP property of COCONUT:
\begin{equation}
p_{BC}=\Upsilon_{p}\frac{\mathbf{B}_{BC}^2}{2}\beta_{\min}+\left(1-\Upsilon_{p}\right)p_{s}
,\end{equation}
where $\Upsilon_p=0.5+0.5\cdot \tanh\left(\frac{\beta_{\min}-\frac{p_s}{0.5 \cdot \mathbf{B}_{BC}^2}}{\beta_{fac}}\cdot \pi\right)$ with $\beta_{fac}=2\times 10^{-6}$ and $\beta_{\min}=0.02$. $\rho_{BC}$ is defined in the same manner as in Eq. (\ref{PPRhoInner}), but with $\rho_{o}$ and $\mathbf{B}$ replaced by $\rho_{s}$ and $\mathbf{B}_{BC}$.

Considering the coronal simulations in this paper are conducted in an inertial coordinate system, we accordingly rotate the GONG-zqs magnetograms to the HCI coordinate system. In quasi-steady-state coronal simulations, the radial component of the inner-boundary magnetic field, $B_{BC,r}$, is linearly interpolated from a time-invariant magnetogram. We use cubic Hermite interpolation for time-evolving coronal simulations, which uses four magnetograms as the stencil. We generate an intermediate magnetogram at each time step and then linearly interpolate $B_{BC,r}$ from this intermediate magnetogram. Meanwhile, the tangential components of the inner-boundary magnetic field, $B_{BC,\theta}$ and $B_{BC,\phi}$, are defined as the corresponding values at the centroid of the nearest inner cell. Although this simplification may reduce accuracy by one order at the inner boundary, the overall accuracy of the numerical solution can still be maintained at the desired order \citep{Gustafsson1981,Wang2016}.
 
Around the 2019 eclipse, the interval between two adjacent GONG-zqs magnetograms is approximately 1 hour, with fluctuations of several minutes. Considering that it is more practical to handle a batch of input magnetogram files with a uniform interval rather than non-uniform ones, we first apply cubic Hermite interpolation to the observed magnetograms to generate a series of magnetograms with a consistent 1-hour interval, which are then used to drive the time-evolving coronal simulations. 
In the time-evolving coronal simulations, step sizes are set to be a few minutes — shorter than the interval between adjacent magnetograms. Therefore, as detailed below, we use cubic Hermite interpolation on the input magnetograms to obtain the required inner-boundary magnetic field for each time step:
\begin{equation}\label{CubicHermitInterpolation}
\begin{aligned}
B_{BC,r}\left(t,\theta,\phi\right)&=h_{00}\left(t^{'}\right)B_r\left(\theta,\phi\right)_m \\
&+ h_{10}\left(t^{'}\right)\left(t_{m+1}-t_m\right)\left(\frac{\partial B_r\left(\theta,\phi\right)}{\partial t}\right)_m \\
&+ h_{01}\left(t^{'}\right)B_r\left(\theta,\phi\right)_{m+1} \\
&+ h_{11}\left(t^{'}\right)\left(t_{m+1}-t_m\right)\left(\frac{\partial B_r\left(\theta,\phi\right)}{\partial t}\right)_{m+1} \\
{\rm with} ~ t^{'}=\frac{t-t_m}{t_{m+1}-t_m}.\
\end{aligned}
\end{equation}
Here $B_{BC,r}\left(t,\theta,\phi\right)$ is the interpolated magnetic field at the inner-boundary facial centroid located at $\left(R_s,\theta,\phi\right)$, $t$ denotes the physical time, and the subscripts `$_m$' and `$_{m+1}$' refer to the m-th and (m+1)-th magnetograms, two adjacent magnetograms nearby $t$, with $t_m<t\leq t_{m+1}$, $\left(\frac{\partial B_r\left(\theta,\phi\right)}{\partial t}\right)_m=\frac{1}{2}\left(\frac{B_r\left(\theta,\phi\right)_{m+1}-B_r\left(\theta,\phi\right)_m}{t_{m+1}-t_m}+\frac{B_r\left(\theta,\phi\right)_{m}-B_r\left(\theta,\phi\right)_{m-1}}{t_{m}-t_{m-1}}\right)$ and $\left(\frac{\partial B_r\left(\theta,\phi\right)}{\partial t}\right)_{m+1}=\frac{1}{2}\left(\frac{B_r\left(\theta,\phi\right)_{m+2}-B_r\left(\theta,\phi\right)_{m+1}}{t_{m+2}-t_{m+1}}+\frac{B_r\left(\theta,\phi\right)_{m+1}-B_r\left(\theta,\phi\right)_{m}}{t_{m+1}-t_{m}}\right)$. Besides, $h_{00}\left(t^{'}\right)=2{t^{'}}^3-3{t^{'}}^2+1$, $h_{10}\left(t^{'}\right)={t^{'}}^3-2{t^{'}}^2+{t^{'}}$, $h_{01}\left(t^{'}\right)=-2{t^{'}}^3+3{t^{'}}^2$ and $h_{11}\left(t^{'}\right)={t^{'}}^3-{t^{'}}^2$ are used to normalize the cubic Hermite interpolation formula.

Additionally, the inner-boundary velocity $\mathbf{v}=\left(u_{BC},v_{BC},w_{BC}\right)$ is defined as the velocity at the centroid of the nearest cell in the inner domain and constrained by a predefined speed $V_{BD, \max}$. Given that an inner-boundary face, a patch of the solar surface, can encompass features like a supergranulation or a sunspot, where the typical average velocity is generally below 1 $\rm {km~s^{-1}}$, we further constrained the facial average plasma velocity at inner-boundary to not significantly exceed this velocity during our simulations:\begin{equation}\label{VInner}
\mathbf{v}_{BC}=\frac{\mathbf{v}_{inner}}{\left|\mathbf{v}_{inner}\right|}\left(\Upsilon_{\mathbf{v}}V_{BD, \max}+\left(1-\Upsilon_{\mathbf{v}}\right)\left|\mathbf{v}_{inner}\right|\right),\
\end{equation}
where $\Upsilon_{\mathbf{v}}=0.5+0.5\cdot \tanh\left(\frac{\left|\mathbf{v}_{inner}\right|-V_{BD, \max}}{V_{BD,fac}}\cdot \pi\right)$ with $V_{BD, \max}=1~\rm {km~s^{-1}}$ and $V_{BD,fac}=2~\rm {m~ s^{-1}}$. Here $\mathbf{v}_{inner}$ refers to the velocity of plasma flow at the centroid of the nearest cell in the inner domain.

The outer boundary of the present model is set to 25$\;R_s$. Since this distance is where the plasma flow has become supersonic {and} super-Alfv{\'e}nic, we prescribed the Neumann boundary conditions at the outer boundary \citep{Brchnelova2022}. Therefore, we extrapolated $r^2B_r$, $B_{\theta}$, $B_{\phi}$, $\rho$, $u$, $v$, $w$ $p,$ and $\psi$ from the outermost cell centres in the computational domain to the ghost cells with a zero gradient \citep{Brchnelova_2022,Perri_2022}.

\section{Numerical results}\label{sec:Numerical Results}
In this section, the time-evolving coronal model developed in this paper is employed to simulate the evolution of the coronal structures during CRs 2219 and 2220, encompassing the solar eclipse on July 2, 2019. About 1300 GONG-zqs magnetograms\footnote{\url{https://gong.nso.edu/data/magmap/QR/zqs/}}, spanning the period from 11:00 on June 29, 2019, to 21:00 on August 22, 2019, are used to drive these simulations.
As mentioned in Sect. \ref{Thecomputationalgridsystemandinitialvaluesetup}, the initial magnetic fields for the quasi-steady-state coronal simulations are achieved from a PF model, with the bottom boundary condition specified by the synoptic map of the radial photospheric magnetic field centred on 11:00 on June 29, 2019. Following the quasi-steady-state coronal simulation, the time-evolving photospheric magnetograms are then used to drive the time-evolving coronal simulations, extending from the solar surface to 25 $R_s$ throughout these two CRs within an inertial coordinate system.}

In Sect.~\ref{sec:TDversusSS}, we perform quasi-steady-state and time-evolving MHD coronal simulations and compare their simulation results to validate the time-evolving approach. It demonstrates that, compared with the quasi-steady-state simulation constrained by only one time-invariant magnetogram during a CR period, there are pronounced differences in the time-evolving simulation results.
In Sect.~\ref{sec:impactoftime stepsizes}, we evaluate the impact of time-step sizes on the time-evolving coronal simulation results. It indicates that using a moderately larger time-step size can significantly enhance computational efficiency with minimal impact on accuracy.

In this paper, all the calculations are performed on the {KU Leuven/UHasselt Tier-2 cluster wICE} of the Vlaams Supercomputer Centrum (VSC)\footnote{\url{https://www.vscentrum.be/}}. Utilising 1,080 CPU cores, the wall-clock time for the time-evolving coronal simulations covering two CRs was approximately 17.54 hours with a 10-minute time-step size and 39.06 hours with a 2-minute time-step size. In the following subsections, we present the results of the MHD coronal simulations for CRs 2219 and 2220.

\subsection{Time-evolving versus quasi-steady-state coronal simulation results}\label{sec:TDversusSS}
In this subsection we compare the time-evolving simulation results at different moments with the quasi-steady-state simulation, which is constrained by a magnetogram at 19:00 on July 2, 2019 — the day of the 2019 eclipse — corresponding to the magnetogram at 82 hours of the time-evolving simulation.
The time-evolving coronal simulation presented here adopts a constant time-step size of 10 minutes. 

\begin{figure}[htpb]
\begin{center}
  \vspace*{0.01\textwidth}
    \includegraphics[width=\linewidth,trim=1 1 1 1, clip]{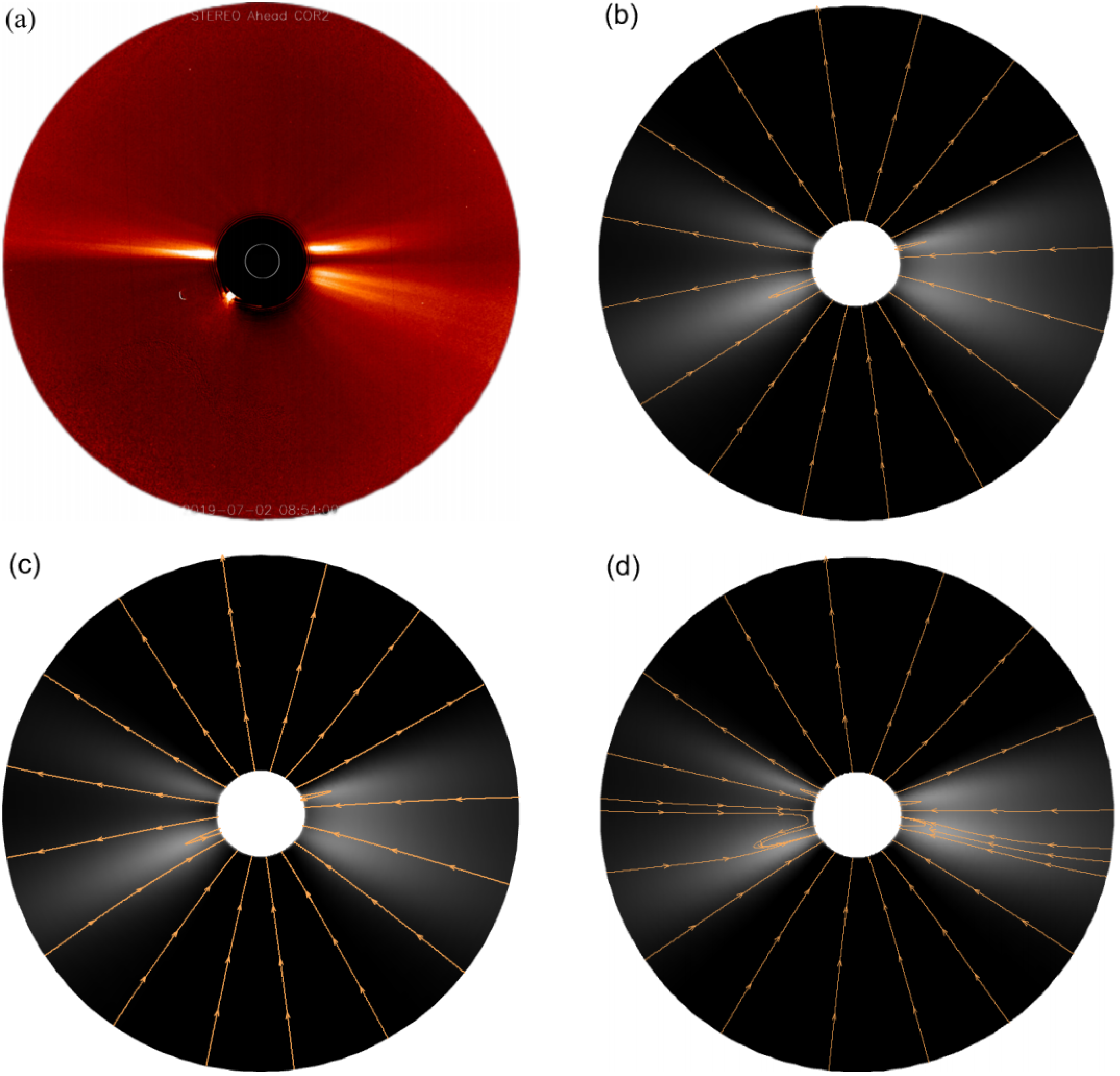}
\end{center}
\caption{White-light {polarisation brightness (pB)} images observed from COR2 (STEREO-A) on July 2, 2019 (a), corresponding pB images synthesised from quasi-steady-state coronal simulation results (b), and synthesised from simulation results at 82 {(c)} and 735 (d) hours of the time-evolving coronal simulations. These synthesised images span from 2.5 to 15$\;R_s$ on the meridian planes perpendicular to the line of sight from STEREO-A. Orange lines highlight magnetic field lines on these selected meridional planes.  The evolution of simulated pB images over this time interval is showcased in online movie 1.}\label{pBMeridianat82and735hrs}
\end{figure}

\begin{figure*}[htpb]
\begin{center}
  \vspace*{0.01\textwidth}
    \includegraphics[width=0.8\linewidth,trim=1 1 1 1, clip]{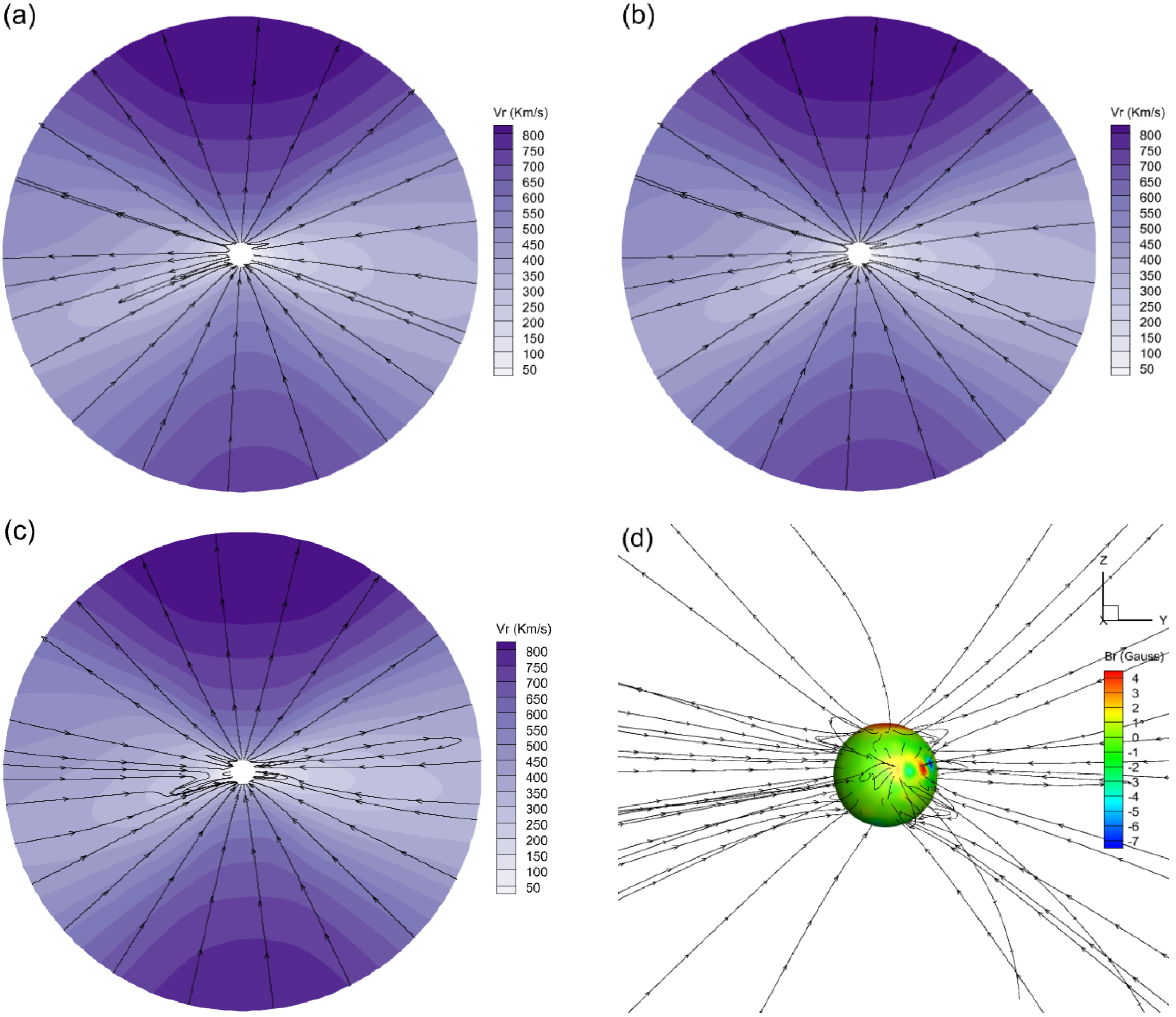}
\end{center}
\caption{2D distribution of magnetic field lines overlaid on contours of the radial plasma speed, $V_r ~\left(\rm km~s^{-1}\right)$, illustrated on meridional planes spanning from 1 to 20 $R_s$ and perpendicular to the STEREO-A line of sight (a, b, and c), and 3D distribution of magnetic field lines (d) at the 82nd hour of the time-evolving simulation. Panel (a) shows the results of the quasi-steady-state simulation, panels (b) and (c) present the time-evolving simulation results at the 82nd and 735th hours, respectively, and panel (d) the 3 D distribution of some selected magnetic field lines. The evolution of these selected magnetic field lines is shown in online movie 2.}\label{VrMeridianat82and735hrs}
\end{figure*}

\begin{figure*}[htpb]
\begin{center}
  \vspace*{0.01\textwidth}
    \includegraphics[width=0.9\linewidth,trim=1 1 1 1, clip]{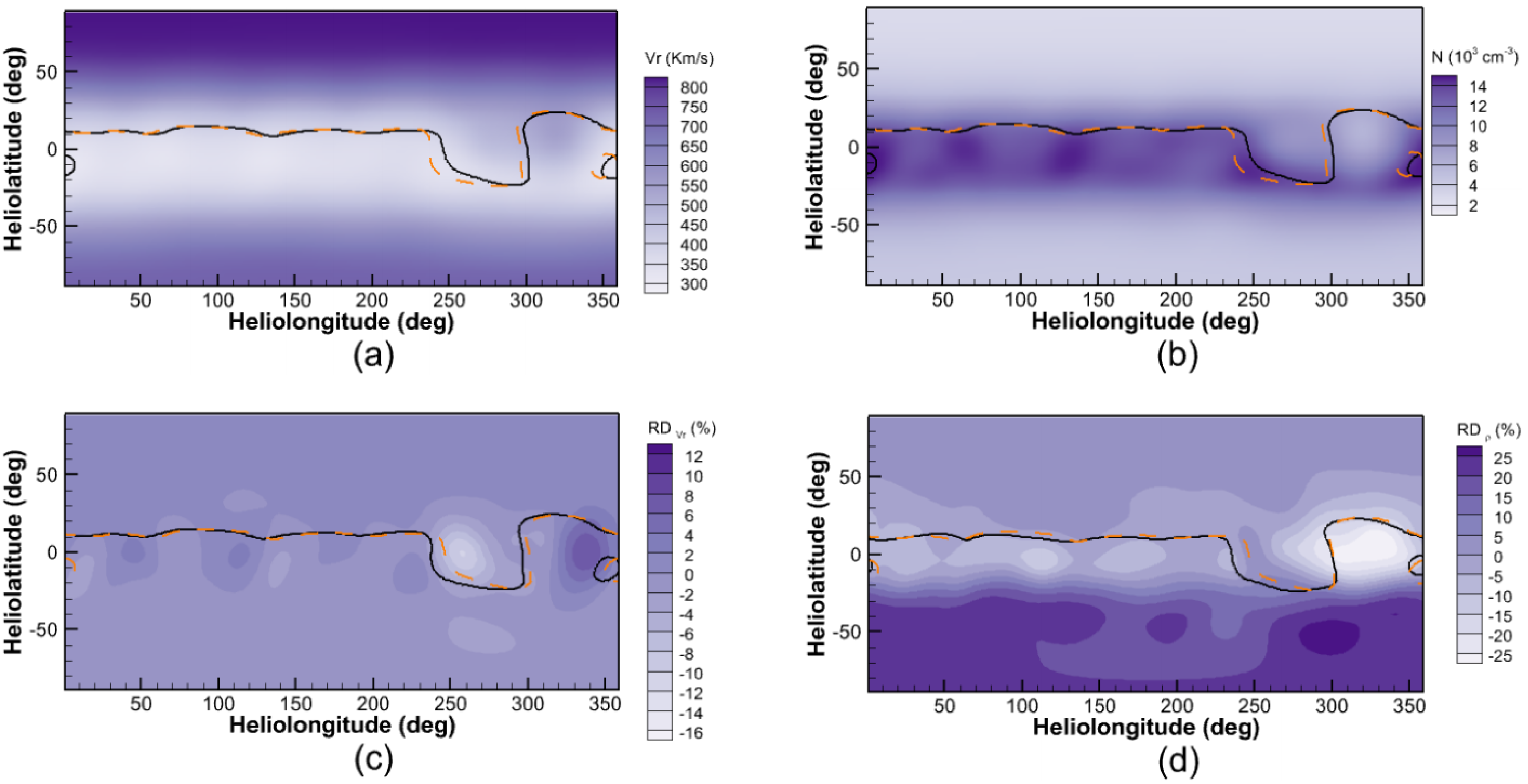}
\end{center}
\caption{Distribution of the radial plasma speeds, $V_r$, in units of $\rm km~s^{-1}$ (a) and proton number density in units of $\rm 10^3 ~ cm^{-3}$ (b) at 20 $R_s$ derived from the 82nd hour of the time-evolving coronal simulation. Panels (c) and (d) show the relative differences in radial velocity and plasma density between the quasi-steady-state and time-evolving coronal simulations. Online movies 3 and 4 illustrate the evolution of these relative differences. The solid black lines and the dashed orange lines represent the MNLs calculated by the time-evolving and quasi-steady-state coronal models.}\label{VrBrat3and20RsPannelat82and735hrs}
\end{figure*}

\begin{figure*}[htpb]
\begin{center}
  \vspace*{0.01\textwidth}
    \includegraphics[width=0.9\linewidth,trim=1 1 1 1, clip]{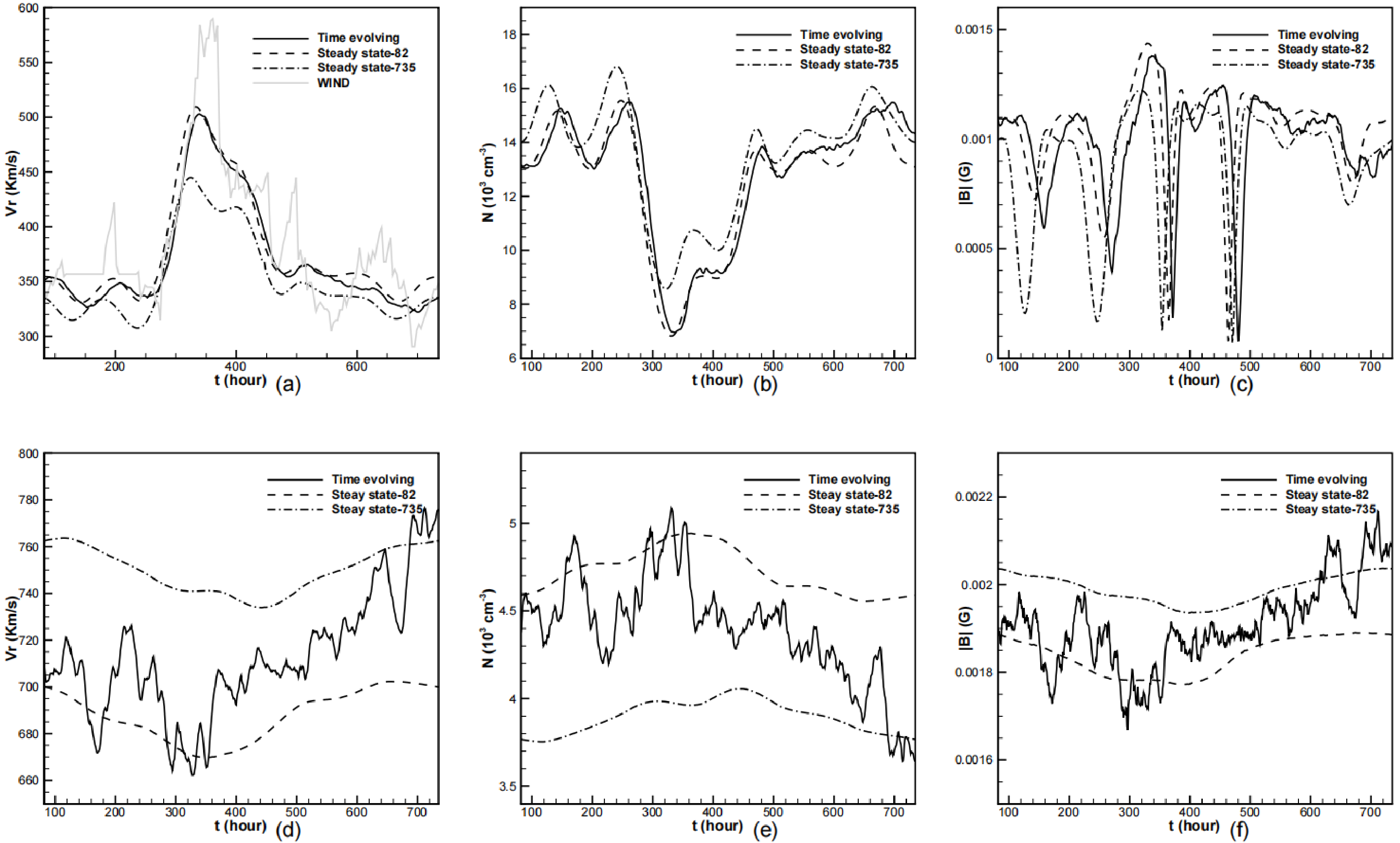}
\end{center}
\caption{Timing diagram of simulated radial velocity, $V_r$, in $\rm km~s^{-1}$ (a, d), proton number density in units of $\rm 10^3 ~ cm^{-3}$ (b, e), and magnetic field strength in unit of Gauss (c, f) observed by two virtual satellites located at the HDLS (a, b, c) and LDHS (d, e, f) regions. The solid black lines denote variables observed at $\left(r, \theta, \phi\right)=\left(20~R_s, 0, 201^{\circ}\right)$ and $\left(r, \theta, \phi\right)=\left(20~R_s, -70^{\circ}, 201^{\circ}\right)$ during the 82nd and 735th hours of the time-evolving coronal simulation. For the quasi-steady-state coronal simulations constrained by magnetograms at 82 (denoted by dashed black lines) and 735 (denoted by dash-dot  black lines) of this time interval, the heliolongitude is mapped to a CR period corresponding to the time series on the horizontal axis of the timing diagram. The solid grey line (a) denotes the velocity observed by the WIND satellite and mapped from 1 AU to 20 $R_s$ following the ballistic propagation.}\label{VrNBtotforHDLS}
\end{figure*}

Movie 2 (linked in Fig.~\ref{VrMeridianat82and735hrs}), spanning from the 82nd to the 735th hour of the time-evolving simulation, depicts the evolution of some selected magnetic field lines within the HCI coordinate system. Movie 1 (linked in {Fig.~\ref{pBMeridianat82and735hrs}}) illustrates the evolution of {polarisation brightness (pB)} images synthesised from the simulation results, as viewed from the perspective of STEREO-A. Furthermore, we transformed the simulation results from the HCI coordinate system into the heliographic co-rotating coordinate system to evaluate the relative differences between the time-evolving and quasi-steady-state simulations. The evolution of these differences is visualised in movies 3 and 4 (linked in Fig.~\ref{VrBrat3and20RsPannelat82and735hrs}). Additionally, in Fig.~\ref{VrNBtotforHDLS}, we present the variation patterns of some simulated plasma parameters in the high-density, low-speed (HDLS) and low-density, high-speed (LDHS) regions. The figure also compares the simulated velocity and observed data, mapped from 1 AU to  20 $R_s$ using ballistic propagation.

In Fig.~\ref{pBMeridianat82and735hrs} we compare white-light pB images from 2.5 to 15 $R_s$ that are observed from the outer coronagraph of the Sun-Earth Connection Coronal and Heliospheric Investigation (SECCHI) instrument suite (panel a) on board the STEREO-A spacecraft \citep{Frazin2012,Howard2008,Kaiser2008TheSM,Thompson2003COR1IC,Thompson2008} and the white-light pB images synthesised from the quasi-steady-state (b) and time-evolving (c and d) coronal simulation results. The top panel demonstrates the pB image observed on July 2, 2019 (a), alongside the image synthesised from the simulation result of quasi-steady-state simulation results (b). The bottom panel presents pB images synthesised from the time-evolving simulation results at the 82nd hour (c) and the 735th hour (d), with an interval of one CR period between the two images. The observed and simulated images reveal four bright structures near the solar equator. However, compared to the observation, the locations of these simulated bright structures in the northern hemisphere are farther north, and the simulated bright structures at the south-east limb extend farther outwards. This discrepancy may be attributed to inaccurate observations for both polar photospheric magnetic fields.

Additionally, it can be seen that when confined by the same magnetogram at the inner boundary, the pB images synthesised from the time-evolving and quasi-steady-state coronal simulation results show only minor differences. After a CR period, the bright structure at the south-west limb shifts about $10^{\circ}$ towards the equator, aligning more closely with the observed results. It demonstrates that bright structures are typically associated with closed-field regions. Additionally, interested readers can refer to online movie 1 to see the evolution of simulated pB images observed from the {COR2 (Coronagraph 2 instrument) on STEREO-A} field of view between the 82nd and 735th hours of the time-evolving coronal simulation.

\begin{figure*}[htpb]
\begin{center}
  \vspace*{0.01\textwidth}
    \includegraphics[width=\linewidth,trim=1 1 1 1, clip]{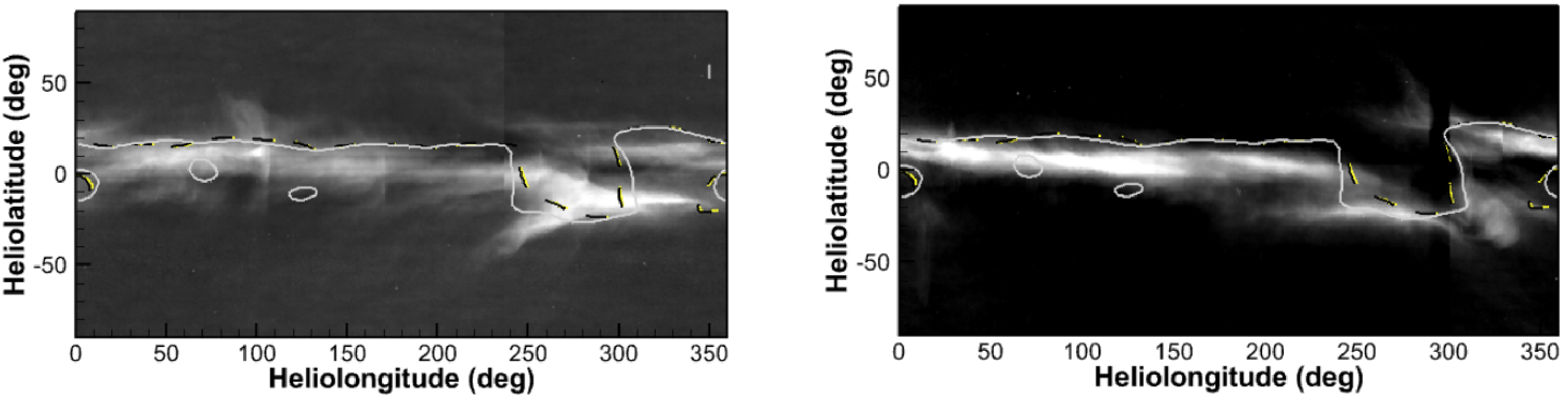}
\end{center}
\caption{Synoptic maps of white-light pB observations from {the SOHO instrument LASCO} C2 at 3$\;R_s$ for CRs 2219 (left) and 2220 (right). The dashed yellow, dashed black, and solid grey lines represent the MNLs derived from the quasi-steady-state simulations and time-evolving simulations at 82nd and 735th hours, respectively.}\label{pBat3Rsat82and735hrs}
\end{figure*}
In Fig.~\ref{pBat3Rsat82and735hrs} we present the synthesised white-light pB images at 3 $R_s$. These images were synthesised from observations during CRs 2219 (panel a) and 2210 (panel b) and displayed in the co-rotating coordinate system overlaid with magnetic field neutral lines (MNLs) from the quasi-steady-state simulation result and from the 82nd and 735th hours of the time-evolving simulation results, respectively. It can be seen that when constrained by the same magnetogram at the inner boundary, the MNLs calculated from the quasi-steady-state and time-evolving simulations are nearly identical. After one CR period, the observed bright structure between $240^{\circ}$ and $310^{\circ}$ in longitude becomes more concave; the MNLs also capture this variation from the time-evolving coronal simulation. Also, it reveals that between $20^{\circ}$ and $240^{\circ}$ in longitude, the segments of these MNLs calculated by MHD models are approximately $10^{\circ}$ farther north than the central axis of the observed bright structures. These discrepancies may be attributed to the empirical, rather than self-consistent, formulation of certain mechanisms in this MHD coronal model — such as coronal heating and solar wind acceleration — which affect magnetic pressure and, consequently, the distribution of MNLs.

In Fig.~\ref{VrMeridianat82and735hrs} we shows the distributions of 2 D magnetic field lines. These radial velocity contours range from 1 to 20 $R_s$ on the same meridians as in Fig.~\ref{pBMeridianat82and735hrs}. The top-left panel (a) denotes the quasi-steady-state simulation result, the top-right and bottom-left panels present the time-evolving simulation results at the 82nd and 735th hours, respectively, and the bottom-right panel (d) demonstrates the 3 D distribution of some selected magnetic field lines. Together with Fig.~\ref{pBMeridianat82and735hrs}, this figure reveals that close-field structures in low-latitude regions dominate the HDLS flow. In contrast, LDHS flows are prevalent in the high-latitude areas, characterised by open-field structures. This is a crucial characteristic of solar minima. 

Besides, it can be seen that after a CR period, the open magnetic field lines in the eastern hemisphere, spanning from  $20^{\circ} S$ to $15^{\circ} N$, change their direction from pointing outwards from the Sun to pointing towards the Sun. Significant variations are also evident in the closed field lines near the Sun. Furthermore, interested readers can refer to online movie 2 to see the evolution of some simulated magnetic field lines in a 3D representation within the HCI coordinate system over the 82nd and 735th hours of the time-evolving coronal simulation. The continuous formation and disappearance of closed field lines, a phenomenon absent in quasi-steady-state coronal simulations, is evident during this time-evolving process.

In Fig.~\ref{VrBrat3and20RsPannelat82and735hrs} we present the distribution of radial velocity (panel a) and proton number density (b) at 20 $R_s$ calculated at the 82nd hour of the time-evolving coronal simulation. We also illustrate relative differences in radial velocity $\rm RD_{V_r}=\frac{V_r^{TE}-V_r^{QSS}}{V_r^{QSS}}$ (c) and plasma density $\rm RD_{\rho}=\frac{\rho^{TE}-\rho^{QSS}}{\rho^{QSS}}$ (d) between the time-evolving and the quasi-steady-state coronal simulations. Furthermore, movies 3 and 4 illustrate the evolution of the relative differences of these parameters. In this comparison, the simulation results at $20 R_s$ during 82 and 735 hours of the time-evolving simulation are compared with the quasi-steady-state coronal simulation constrained by a time-invariant magnetogram. It can be seen that the relative differences in radial velocity $\rm RD_{V_r}$ and plasma density $\rm RD_{\rho}$ can reach $15\%$  and $25\%$. Additionally, in Table \ref{QSSVSTE} we list the average relative differences in plasma density ${\rm RD}_{{\rm ave},\rho}$, velocity ${\rm RD}_{{\rm ave},\left|\mathbf{v}\right|}$, and magnetic field strength ${\rm RD}_{{\rm ave},\left|\mathbf{B}\right|}$ between the steady-state and the time-evolving coronal simulation results at the 82nd and 735th hours, respectively.
\begin{table}
\centering
\caption{Average relative differences between quasi-steady-state and time-evolving coronal simulation results.}
\label{QSSVSTE}
\begin{tabular}{llll}
\hline\noalign{\smallskip}
 Parameters & ${\rm RD}^t_{{\rm ave},\rho}$  & ${\rm RD}^t_{{\rm ave},\left|\mathbf{v}\right|}$ & ${\rm RD}^t_{{\rm ave},\left|\mathbf{B}\right|}$ \\
\noalign{\smallskip}\hline\noalign{\smallskip}
 $t$=82~hrs & $0.54\%$  & $1.06\%$  & $4.24\%$ \\
 $t$=735~hrs & $4.53\%$  & $8.49\%$  & $35.35\%$ \\
\noalign{\smallskip}\hline
\end{tabular}
\end{table}
Here
$${\rm RD}_{{\rm ave},\rho}=\frac{\sum\limits_{i=1}^N \big|\rho_{i, {\rm QSS}}-\rho_{i,{\rm TE}} \big|}{\sum\limits_{i=1}^N \rho_{i, {\rm QSS}}},\ $$
$${\rm RD}_{{\rm ave},\left|\mathbf{v}\right|}=\frac{\sum\limits_{i=1}^N \big|\mathbf{v}_{i, {\rm QSS}}-\mathbf{v}_{i,{\rm TE}} \big|}{\sum\limits_{i=1}^N \left|\mathbf{v}_{i, {\rm QSS}}\right|},\ $$
$${\rm RD}_{{\rm ave},\left|\mathbf{B}\right|}=\frac{\sum\limits_{i=1}^N \big|\mathbf{B}_{i, {\rm QSS}}-\mathbf{B}_{i,{\rm TE}} \big|}{\sum\limits_{i=1}^N \left|\mathbf{B}_{i, {\rm QSS}}\right|}.\ $$
The superscript `$^{t}$' means the $t$-th hour of the time-evolving simulation, the subscript `$_{i,{\rm QSS}}$' or `$_{i,{\rm TE}}$' denotes the corresponding variable in cell $i$ calculated by the quasi-steady-state or time-evolving coronal models, and $N$ is the number of cells in the computational domain. It can be seen that the overall differences in the simulation plasma densities, velocities, and magnetic field strength between the quasi-steady-state and time-evolving coronal models are relatively small when constrained by the same magnetogram. However, these differences can increase significantly during a CR period, particularly in magnetic field strength, reaching up to $35\%$.

In Fig.~\ref{VrNBtotforHDLS} we display the timing diagrams of radial velocity (panels a and d), proton number density (b and e) and magnetic field strength (c and f) during the 82nd and 735th hours of the time-evolving coronal simulation by solid black lines. These parameters are observed by two virtual satellites placed at $\left(r, \theta, \phi\right)=\left(20~R_s, 0, 201^{\circ}\right)$ and $\left(r, \theta, \phi\right)=\left(20~R_s, -70^{\circ}, 201^{\circ}\right)$, which located in the HDLS (q, b, c) and LDHS (d, e, f) regions, respectively. Additionally, we mapped the heliolongitude in quasi-steady-state coronal simulations to a CR period of physical time as
\begin{equation*}
t=\left\{\begin{array}{c}
82-653\frac{\phi-\phi_0}{360^{\circ}},  \text { if } \phi \leq \phi_0 \\
735-653\frac{\phi-\phi_0}{360^{\circ}},  \text { if } \phi > \phi_0
\end{array}\right.,\
\end{equation*}
where $\phi_0=201^{\circ}$. 
The selected variables were calculated using the quasi-steady coronal models, constrained by magnetograms at 19:00 on July 2, 2019, and 00:00 on July 30, 2019, along the mapped time at $\left(r, \theta\right)=\left(20~R_s, 0\right)$. It reveals that the peaks and troughs in the timing diagrams from the time-evolving simulation differ from those in the quasi-steady-state coronal simulations, with these differences being more pronounced in the LDHS regions than in the HDLS regions. In the LDHS regions, the peaks and troughs observed around the 160th, 220th, 320th, and 660th hours in the time-evolving coronal model are absent in the quasi-steady-state simulations. Additionally, in the HDLS regions, the troughs in the magnetic field strength timing diagrams occur earlier in the time-evolving simulation than those from the quasi-steady-state models.

Furthermore, we made a ballistic propagation to map the observed velocity at 1 AU around the equatorial plane to 20$\;R_s$ and compared them with the simulation results, as shown in the left-top panel of Fig.~\ref{VrNBtotforHDLS}. Both quasi-steady and time-evolving simulations capture the velocity peak between 250 and 490 hours of the coronal simulation. However, between 300 and 330 hours and 370 and 470 hours, the time-evolving simulation result aligns more closely with the observation. It shows that the time-evolving coronal model has the potential to make simulation results coincide more closely with observations.

\subsection{The impact of time-step sizes on simulation results}\label{sec:impactoftime stepsizes}
In this subsection we evaluate the impact of time-step sizes on the time-evolving coronal simulation results. The model performs well with a time-step size of 10 minutes or less. However, increasing the time steps to larger intervals, such as 20 minutes, can cause the simulation to crash under certain conditions. 
To further evaluate the impact of time-step sizes on accuracy and efficiency, we conducted a time-evolving coronal simulation with a smaller time-step size of 2 minutes and compared the simulated plasma density, velocity, and magnetic field strength with those from the previously discussed simulation, which used a 10-minute time step.

In the time-evolving coronal simulation with time steps of 10 minutes, approximately 7 to 9 Newtonian iterations, as described in Sect. (\ref{discretization}), are typically performed in each time step. However, instances exist where the $L_2$ norm of the differences between the state variables updated in two consecutive Newton iterations fluctuates around a value larger than the threshold used to stop the Newtonian iterations. In such cases, we terminated the Newton iterations at the tenth iteration. For the time-evolving simulation with 2-minute time steps, approximately five to six Newtonian iterations are typically performed per time step; the computational time was about 2.23 times longer than in the simulation with a 10-minute time step.

\begin{figure*}[htpb]
\begin{center}
  \vspace*{0.01\textwidth}
    \includegraphics[width=0.8\linewidth,trim=1 1 1 1, clip]{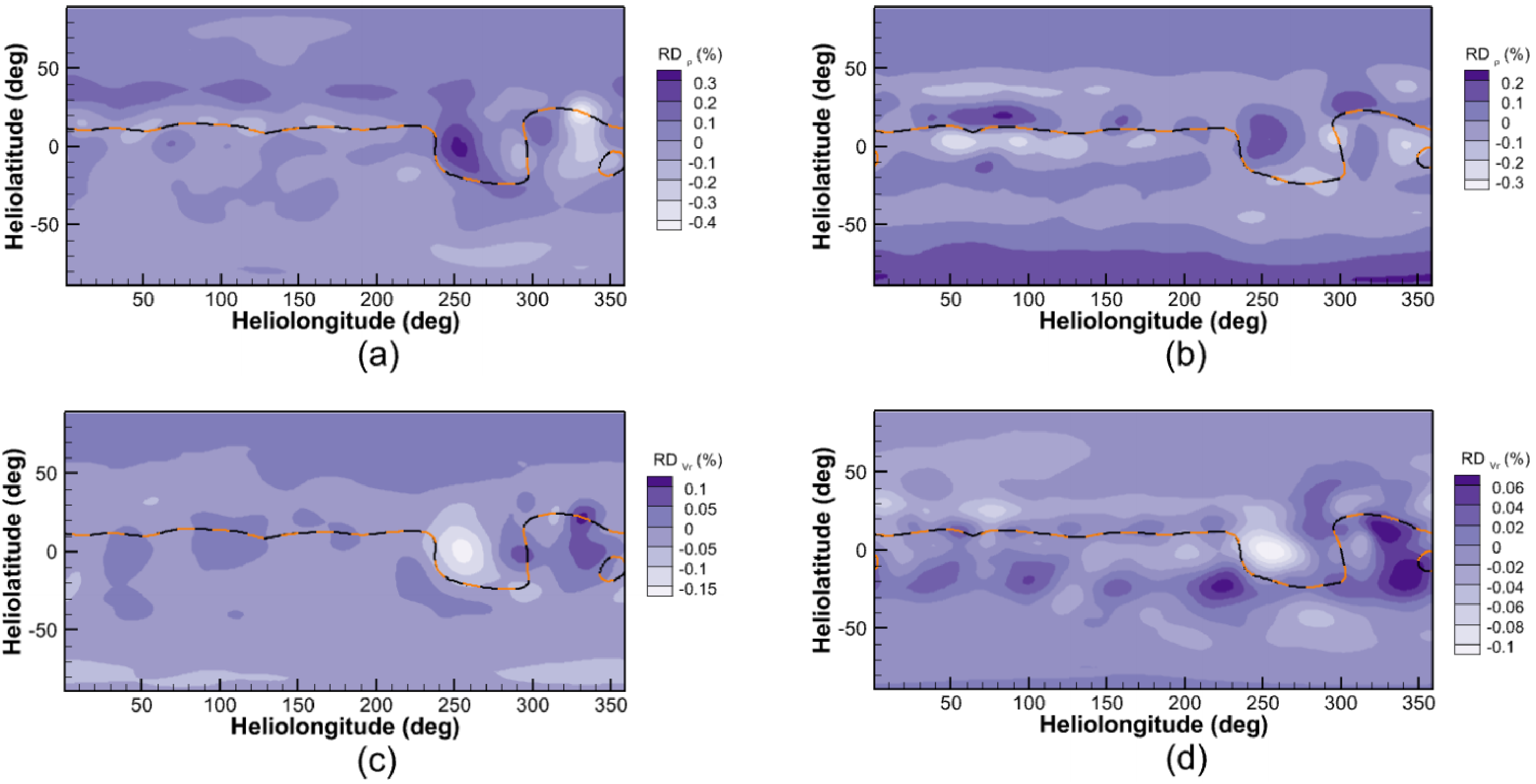}
\end{center}
\caption{Distribution of the relative differences in plasma density (a, b) and radial velocity (c, d) at 20 $R_s$ between the results calculated with time steps of 2 and 10 minutes. The solid black and dashed orange lines denote the MNLs calculated with time steps of 2 and 10 minutes, respectively. The left and right panels denote the results at the 82nd and 735th hour of the time-evolving coronal simulations.}\label{ADrhoandADVrat20RsPannelat82and735hrs}
\end{figure*}
In Fig.~\ref{ADrhoandADVrat20RsPannelat82and735hrs} we show the relative differences in plasma density $\rm RD_{\rho}$ (panels a and b) and radial velocity $\rm RD_{V_r}$ (c and d) between the results calculated with time steps of 2 and 10 minutes at 20 $R_s$. Here, $\rm RD_{\rho}$ and $\rm RD_{V_r}$ are defined similar to those in Fig.~\ref{VrBrat3and20RsPannelat82and735hrs}, but with the variables from the quasi-steady-state simulation replaced by their counterparts from the time-evolving simulation with a time step of 2 minutes. Overlaid on these contours are the MNLs calculated with time steps of 2 and 10 minutes. It can be seen that the MNLs calculated with different time-step sizes coincide with each other. In most regions, the relative differences in plasma density and radial velocity are less than $0.2 \%$ and $0.1 \%$. Additionally, it reveals that the most significant relative differences mainly appear around the MNLs.

\begin{table*}
\caption{Comparison of time-evolving coronal simulations with $\rm dt$ = 10 and 2 minutes for two CRs of physical time.}
\label{largeDtVSsmallDt}
\begin{tabular}{llll}
\hline\noalign{\smallskip}
wall-clock times for $\rm dt$=2 min \& $\rm dt$=10 min (hrs) & ${\rm RD}_{{\rm ave},\rho}^{82~ {\rm hrs}}$  \& ${\rm RD}_{{\rm ave},\rho}^{735~ {\rm hrs}}$ & ${\rm RD}_{{\rm ave},\left|\mathbf{v}\right|}^{82~ {\rm hrs}}$  \& ${\rm RD}_{{\rm ave},\left|\mathbf{v}\right|}^{735~ {\rm hrs}}$ & ${\rm RD}_{{\rm ave},\left|\mathbf{B}\right|}^{82~ {\rm hrs}}$  \& ${\rm RD}_{{\rm ave},\left|\mathbf{B}\right|}^{735~ {\rm hrs}}$\\
\noalign{\smallskip}\hline\noalign{\smallskip}
 39.06 \& 17.54 & $0.09\%$ \& $0.08\%$ & $0.10\%$ \& $0.09\%$ & $0.60\%$ \& $0.42\%$\\
\noalign{\smallskip}\hline
\end{tabular}
\end{table*}
Furthermore, Table \ref{largeDtVSsmallDt} presents the average relative differences in plasma density ${\rm RD}_{{\rm ave},\rho}^{t}$, velocity ${\rm RD}_{{\rm ave},\left|\mathbf{v}\right|}^{t}$ and magnetic field ${\rm RD}_{{\rm ave},\left|\mathbf{B}\right|}^{t}$ between the simulation results calculated with time steps of 2 and 10 minutes. Here, ${\rm RD}_{{\rm ave},\rho}^{t}$, ${\rm RD}_{{\rm ave},\left|\mathbf{v}\right|}^{t}$ and ${\rm RD}_{{\rm ave},\left|\mathbf{B}\right|}^{t}$ are defined similarly to those in Table \ref{QSSVSTE}, except that the variables from the quasi-steady-state simulation are replaced by their counterparts from the time-evolving simulation with a time-step size of 2 minutes. Additionally, the table includes the wall-clock time required for both time-evolving coronal simulations.
It indicates that the overall differences between the simulation results with $\rm dt=10~min$ and $\rm dt=2~min$ are minimal, remaining below $1\%$. In contrast, the computational time with $\rm dt=2~min$ is approximately 2.23 times as long as that with $\rm dt=10~min$. This demonstrates that the computational efficiency of the implicit time-evolving coronal model can be significantly enhanced by increasing the time-step size while still maintaining the required level of temporal accuracy.

\section{Concluding remarks}\label{sec:Conclusion}
The commonly used steady-state or time-dependent coronal models are typically constrained by a time-invariant magnetogram, which fails to reflect that the corona evolves. This simplification usually discards crucial information needed to simulate solar disturbances accurately, such as CME propagation. It also leads to discrepancies compared with 1 AU observations, resulting in an overly simplistic representation of magnetic connectivity. In contrast, time-evolving coronal models, driven by a series of more realistic time-evolving magnetograms, can capture the dynamic features of the corona with higher fidelity, offering the potential to address these shortcomings.

In this study we extended the recently developed COCONUT coronal model, constrained initially by a time-invariant magnetogram, to a more realistic time-evolving coronal model that is driven by a sequence of time-varying magnetograms, enabling it to capture the dynamic features of the solar corona. It is the first fully implicit time-evolving MHD coronal model, offering the flexibility to use large time-step sizes exceeding the CFL condition. This paper demonstrates that accurate real-time simulations of the time-evolving global corona can be achieved using no more than 20 CPU cores.  This approach is considerably more efficient than explicit or semi-implicit coronal models. 

A series of hourly updated GONG-zqs magnetograms around the 2019 eclipse drive the time-evolving coronal simulations in an inertial coordinate system. The cubic Hermite interpolation method reconstructs the inner-boundary magnetic field at each physical time step. Furthermore, we designed an HLL Riemann solver with a self-adjustable dissipation term, which reduces the dissipation term for low-speed flows and reverts to the original one for high-speed flows. This method minimises the degradation of the solution accuracy in low-speed regions and improves the numerical stability of our coronal model. Additionally, during the time-evolving coronal simulations, the model adaptively adjusts the plasma density to avoid the abnormally high Alfv{\'e}nic speed. This strategy enhances the PP property of the present coronal model. 

The simulation results demonstrate that this implicit time-evolving coronal model successfully replicates observed coronal structures, including those captured in pB images and in situ observations. Additionally, the comparison with a quasi-steady-state coronal simulation, constrained by a single time-invariant magnetogram during a CR period, demonstrates that the time-evolving coronal model with a 10-minute time step can produce coronal structures comparable to those generated by the quasi-steady-state model at the moment when both models use the same magnetogram to define their inner boundaries. The average relative differences in plasma density, velocity, and magnetic field strength between the time-evolving and quasi-steady-state coronal simulations are less than $0.6 \%$, $1.3 \%,$ and $4.3 \%$, respectively.
However, these differences can increase significantly during a CR period, with the magnetic field strength differences reaching as high as $35\%$. Consequently, we can conclude that the time-evolving coronal model with a significant time-step size not only reproduces coronal structures comparable to those generated by the quasi-steady-state model at the moment confined by the same magnetogram at the inner boundary; it also captures the continuous temporal evolution of these structures, a capability absent in the quasi-steady-state model.

Additionally, we evaluated the impact of time-step size on the simulation results. The average relative differences in plasma density, velocity, and magnetic field strength between simulations with time steps of 2 minutes and 10 minutes are no more than $0.09 \%$, $0.10 \%$, and $0.60 \%$, respectively. Meanwhile, the computational time required for a 10-minute time-step size is less than half of that needed for a 2-minute time-step size. We can conclude that increasing the time-step size can significantly enhance the computational efficiency of the fully implicit time-evolving coronal model while still preserving the required temporal accuracy. We also notice that excessively large time-step sizes, such as 20 minutes, can occasionally cause the code to crash. 
This time-evolving coronal model is designed to be coupled with an inner heliosphere model in the future. The implicit model's flexibility in selecting significant time steps will allow us to determine the time steps of the MHD coronal model via the inner heliosphere model. It is well known that most MHD inner heliosphere models use explicit numerical schemes, with time steps limited by the CFL stability criterion to around 10 minutes. Thus, a 10-minute time step is efficient and numerically stable enough for our research.

Given that the computational time of 1281 hours of physical time is no more than 18 hours when adopting a time-step size of 10 minutes, it is practical to adopt a smaller {time step} to achieve more accurate simulation results for higher-frequency events such as CMEs, albeit at the cost of a manageable reduction in computational efficiency. The physical time-step size can be adjusted based on the accuracy required for specific research or applications. This allows for optimised time-step sizes that balance computational efficiency with the necessary numerical stability and accuracy. This approach enabled us to model the entire solar-terrestrial domain with a single model, thereby streamlining the Sun-to-Earth forecasting process.

Although this established fully implicit time-evolving MHD solar coronal model has many advantages and is a promising tool for timely and accurately simulating the time-evolving corona in practical space weather forecasting, there remains considerable room for improvement.
In future work, we plan to couple this highly efficient time-evolving coronal model with an inner heliosphere model, such as EUHFORIA, to further validate its potential for enhancing the performance of the current Sun-to-Earth model chain in providing timely and accurate space weather forecasting. Additionally, we plan to incorporate surface flux transport models, such as the advective flux transport model \citep{Upton2014a}, to advect the radial magnetic field with observed flows. This approach may help generate a more realistic magnetic field evolution at the inner boundary of our coronal model. Furthermore, we may also try to incorporate additional observations into the inner boundary conditions, such as inferring horizontal velocities from observational data using the time-distance helioseismology method \citep{Yalim_2017,Zhao2012}. Additionally, reducing magnetic field discretisation errors by solving decomposed MHD equations \cite[e.g.][]{Feng_2010,Licaixia2018,WANG201967} could enhance the numerical stability of COCONUT, even when addressing extremely low-$\beta$ problems.

\begin{acknowledgements}
This project has received funding from the European Research Council Executive Agency (ERCEA) under the ERC-AdG agreement No. 101141362 (Open SESAME). 
These results were also obtained in the framework of the projects FA9550-18-1-0093 (AFOSR), C16/24/010  (C1 project Internal Funds KU Leuven), G0B5823N and G002523N (WEAVE) (FWO-Vlaanderen), and 4000145223 (SIDC Data Exploitation (SIDEX), ESA Prodex).

This work is also part of a project supported by the Specialized Research Fund for State Key Laboratories, managed by the Chinese State Key Laboratory of Space Weather. It is also supported by the National Natural Science Foundation of China (grant Nos. 42030204, 42074208 and 42104168).

The resources and
services used in this work were provided by the VSC (Flemish Supercomputer Centre), funded by the Research Foundation – Flanders (FWO) and the Flemish Government. The Research Council of Norway supports F.Z.\ through its Centres of Excellence scheme, project No. 262622.
This work utilises data obtained by the Global Oscillation Network Group (GONG) program, managed by the National Solar Observatory and operated by AURA, Inc., under a cooperative agreement with the National Science Foundation. The data were acquired by instruments operated by the Big Bear Solar Observatory, High Altitude Observatory, Learmonth Solar Observatory, Udaipur Solar Observatory, Instituto de Astrof{\'i}sica de Canarias, and Cerro Tololo Inter-American Observatory. The authors also acknowledge the use of the STEREO/SECCHI data produced by a consortium of the NRL (US), LMSAL (US), NASA/GSFC (US), RAL (UK), UBHAM (UK), MPS (Germany), CSL (Belgium), IOTA (France), and IAS (France). 
\end{acknowledgements}

\bibliographystyle{aa}
\bibliography{COCONUT}
\clearpage

\begin{appendix}
\section{HLL Riemann solver with a self-adjustable dissipation term }\label{HLLsad}
Approximate Riemann solvers, such as HLL- \citep{Feng_2021} and AUSM-type \citep{Kitamura2020chap3,Wang_2022} solvers, can be used to compute the inviscid flux $\mathbf{F}_{n,ij}$ between two cells $i$ and $j$. The subscript $_{i,j}$ is omitted hereafter for simplicity. In this study we adopted the HLL Riemann solver, which can be described as follows:
\begin{equation}\label{numericalflux}
\mathbf{F}_{n}\left(\mathbf{U}_{L},\mathbf{U}_{R}\right)=\frac{\mathbf{F}_n\left(\mathbf{U}_{L}\right)+\mathbf{F}_n\left(\mathbf{U}_{R}\right)}{2}-\frac{1}{2}\mathbf{D}_{\rm HLL}\left(\mathbf{U}_{L},\mathbf{U}_{R}\right),
\end{equation}
with
\begin{equation}\label{InviscidFlux}
\begin{aligned}
\mathbf{D}_{\rm HLL}\left(\mathbf{U}_{L},\mathbf{U}_{R}\right)=&\frac{\left(S_L+S_R\right)\left(\mathbf{F}_{n}\left(\mathbf{U}_{R}\right)-\mathbf{F}_{n}\left(\mathbf{U}_{L}\right)\right)}{S_R-S_L}\\
&-\frac{2S_RS_L}{S_R-S_L}\left(\mathbf{U}_{R}-\mathbf{U}_{L}\right),
\end{aligned}
\end{equation}
and
\begin{equation*}
    \mathbf{F}_n\left(\mathbf{U}\right)=
                     \begin{pmatrix}
                        V_{n}\rho  \\
                         V_{n}\rho u+p_T n_{x}-B_nB_x\\
                         V_{n}\rho v +p_T n_{y}-B_nB_y\\
                         V_{n}\rho w +p_T n_{z}-B_nB_z\\\
                        \left(v B_x-B_y u\right) n_{y}+\left(w B_x-B_z u\right) n_{z} +\psi n_{x}\\
                        \left(u B_y-B_x v\right) n_{x}+\left(w B_y-B_z v\right) n_{z} +\psi n_{y}\\
                        \left(u B_z-B_x w\right) n_{x}+\left(v B_z-B_y w\right) n_{y} +\psi n_{z}\\
                        V_n \left(E+p_T\right)-B_n\mathbf{B}\cdot\mathbf{v}\\
                        B_n V_{\rm ref}^2
                     \end{pmatrix}.\
\end{equation*}
Here, $V_n\left(\mathbf{U}\right)=\mathbf{v} \cdot \mathbf{n}_{ij}$ and $B_n\left(\mathbf{B}\right)=\mathbf{B} \cdot \mathbf{n}_{ij}$ denote the velocity and magnetic field along the normal direction of $\Gamma_{ij}$, $S_{L}$ and $S_{R}$ usually stand for the conventional two fast waves in the HLL Riemann solver \citep{EINFELDT1991273}. Unless otherwise stated, the subscripts `$_{L}$'  and `$_{R}$' denote the corresponding left and right variables evaluated on the centroid of $\Gamma_{ij}$. In this paper, we set $S_{L}=\min \left(0, \lambda_{\min}\left(\mathbf{U}_L\right),\lambda_{\min}\left(\mathbf{U}_R\right)\right)$ and $S_{R}=\max \left(0, \lambda_{\max}\left(\mathbf{U}_L\right),\lambda_{\max}\left(\mathbf{U}_R\right)\right)$ with $$\lambda_{\min}\left(\mathbf{U}\right)=\min\left(V_n\left(\mathbf{U}\right),V_n\left(\mathbf{U}\right)\pm c_{f~{\rm or}~s}\left(\mathbf{U}\right),V_n\left(\mathbf{U}\right)\pm c_A\left(\mathbf{U}\right)\right),$$ and $$\lambda_{\max}\left(\mathbf{U}\right)=\max\left(V_n\left(\mathbf{U}\right),V_n\left(\mathbf{U}\right)\pm c_{f~{\rm or}~s}\left(\mathbf{U}\right),V_n\left(\mathbf{U}\right)\pm c_A\left(\mathbf{U}\right)\right),$$
where $c_{f~{\rm or}~s}\left(\mathbf{U}\right)=\sqrt{0.5\left(\frac{\gamma p}{\rho}+\frac{\mathbf{B}^2}{\rho} \pm \sqrt{\left(\frac{\gamma p}{\rho}+\frac{\mathbf{B}^2}{\rho}\right)^2-4\frac{\gamma p}{\rho}\frac{B_{n}^2}{\rho}}\right)}$, and $c_A\left(\mathbf{U}\right)=\frac{\left|B_{n}\right|}{\rho^{0.5}}$.

Although the HLL Riemann solver performs well in preserving the PP property for compressible high-speed flows with large Mach numbers, its diffusion is excessive for incompressible low-speed flows with small Mach numbers. Due to excessive diffusion in the original solver, the accumulation of solution accuracy degradation in low-speed flow regions may eventually lead to code crashes. A suitable low-dissipation numerical flux solver is required for low-speed flows to avoid a degradation of the solution accuracy \cite[e.g.][]{Esquivel2010,Kitamura2011,Kitamura2018,Minoshima_2020,MINOSHIMA2021110639,Wang_2022}. Therefore, we introduced the factor
\begin{equation}
\varphi=\frac{\max \left( \left|S_L \right|,\left|S_R \right| \right)} {S_R-S_L}
\end{equation}
to the second term on the right-hand side  of Eq. (\ref{InviscidFlux}), which plays a major role in low Mach number regions and decreases to zero in high Mach number regions to reduce the diffusion of HLL Riemann solver for low-speed flow. Consequently, we get the following HLL Riemann solver,
\begin{equation}\label{numericalfluxSelfAdjustable}
\mathbf{F}_{n,ij}\left(\mathbf{U}_{L},\mathbf{U}_{R}\right)=\frac{\mathbf{F}_n\left(\mathbf{U}_{L}\right)+\mathbf{F}_n\left(\mathbf{U}_{R}\right)}{2}-\frac{1}{2}\mathbf{D}_{\rm HLL}^{'}\left(\mathbf{U}_{L},\mathbf{U}_{R}\right),
\end{equation}
where
\begin{equation}\label{InviscidFluxSelfAdjustable}
 \begin{aligned}
\mathbf{D}_{\rm HLL}^{'}\left(\mathbf{U}_{L},\mathbf{U}_{R}\right)=&\frac{\left(S_L+S_R\right)\left(\mathbf{F}_{n}\left(\mathbf{U}_{R}\right)-\mathbf{F}_{n}\left(\mathbf{U}_{L}\right)\right)}{S_R-S_L}\\
&-\varphi\frac{2S_RS_L}{S_R-S_L}\left(\mathbf{U}_{R}-\mathbf{U}_{L}\right).
 \end{aligned}
\end{equation}
The factor $\varphi$ will reduce the dissipation term of the HLL Riemann solver by half for low-speed flows and recover to the original one for high-speed flows. We find that incorporating this factor enhances the numerical stability of our coronal model, as the model would be prone to crashing without this adjustment.

\end{appendix}
\end{document}